\documentclass[fleqn,usenatbib,usedcolumn]{mnras}
\usepackage[british]{babel}             
\usepackage{ae,aecompl}
\usepackage{newtxtext}                  
\usepackage[slantedGreek]{newtxmath}    
\usepackage[T1]{fontenc}                
\usepackage{graphicx}                   
\usepackage{amsmath}    
\usepackage{amssymb}    
\usepackage{color} 
\usepackage{hyperref}
\newcommand{\Msun}{\hbox{$\rm\thinspace M_{\odot}$}}

\newcommand{\irx}{\mbox{\rm IRX}}

\title[Dust attenuation in $2<z<3$ star-forming galaxies]{Dust attenuation
  in $\bmath {2<z<3}$ star-forming galaxies from deep ALMA observations of the
\textbf{\textit{Hubble Ultra Deep Field}}}
\author [R.J. McLure et al.]{R. J. McLure$^{1}$\thanks{E-mail:  rjm@roe.ac.uk},
J.S. Dunlop$^{1}$, F. Cullen$^{1}$, N. Bourne$^{1}$, P.N. Best$^{1}$, S. Khochfar$^{1}$, 
\newauthor R.A.A. Bowler$^{2}$, A.D.~Biggs$^{3}$,
J.E.~Geach$^{4}$, D. Scott$^{5}$, 
M.J.~Micha{\l}owski$^{6,1}$, \newauthor W. Rujopakarn$^{7,8,9}$, E. van Kampen$^{3}$, A. Kirkpatrick$^{10}$, A. Pope$^{11}$\\
$^{1}$Institute for Astronomy, University of Edinburgh, Royal Observatory, Edinburgh, EH9 3HJ, UK\\
$^{2}$Astrophysics, The Denys Wilkinson Building, University of Oxford, Keble Road, Oxford, OX1 3RH, UK\\
$^{3}$European Southern Observatory, Karl-Schwarzchild-Str. 2, 95748 Garching b. Munchen, Germany\\
$^{4}$Centre for Astrophysics Research, Science \& Technology Research
Institute, University of Hertfordshire, Hatfield, AL10 9AB, UK\\
$^{5}$Department of Physics and Astronomy, University of British Columbia, Vancouver, BC V6T1Z1, Canada\\
$^{6}$Astronomical Observatory Institute, Faculty of Physics, Adam Mickiewicz University, ul.~S{\l}oneczna 36, 60-286 Pozna{\'n}, Poland\\
$^{7}$Department of Physics, Faculty of Science, Chulalongkorn University, 254 Phayathai Road, Pathumwan, Bangkok 10330, Thailand\\
$^{8}$National Astronomical Research Institute of Thailand (Public Organization), Donkaew, Maerim, Chiangmai 50180, Thailand\\
$^{9}$Kavli Institute for the Physics and Mathematics of the Universe (WPI), The University of Tokyo, Kashiwa, Chiba 277-8583, Japan\\
$^{10}$Yale Center for Astronomy \& Astrophysics, Physics Department, P.O. Box 208120, New Haven, CT 06520, USA\\
$^{11}$Department of Astronomy, University of Massachusetts, Amherst, MA 01002, USA}
\date{Accepted XXX. Received YYY; in original form ZZZ}

\pubyear{2017}

\begin{document}
\label{firstpage}
\pagerange{\pageref{firstpage}--\pageref{lastpage}}
\maketitle

\begin{abstract}
We present the results of a new study of the relationship between infrared
excess (IRX$\equiv L_{\rm IR}/L_{\rm UV}$), UV spectral slope ($\beta$) and stellar
mass at redshifts $2<z<3$, based on a deep Atacama Large Millimeter Array (ALMA)
1.3-mm continuum mosaic of the {\it Hubble Ultra Deep Field} (HUDF). 
Excluding the most heavily-obscured sources, we use a stacking analysis
to show that $z\simeq 2.5$ star-forming galaxies in the mass range
$9.25\leq \log(M_{\ast}/\Msun) \leq 10.75$ are fully consistent with 
the IRX$-\beta$ relation expected for a relatively grey attenuation
curve, similar to the commonly adopted Calzetti law. Based on a large,
mass complete, sample of $2\leq z \leq 3$ star-forming galaxies 
drawn from multiple surveys, we proceed to
derive a new empirical relationship between $\beta$ and stellar mass,
making it possible to predict UV attenuation ($A_{1600}$) and IRX
as a function of stellar mass, for any assumed attenuation law. 
Once again, we find that $z\simeq 2.5$ star-forming galaxies follow
$A_{1600}$--$M_{\ast}$ and IRX$-M_{\ast}$ relations consistent with a
relatively grey attenuation law, and find no compelling evidence that
star-forming galaxies at this epoch follow a reddening law as steep as
the Small Magellanic Cloud (SMC) extinction curve. In fact, we use a simple simulation to
demonstrate that previous determinations of the IRX$-\beta$ 
relation may have been biased toward low values of IRX at red values of
$\beta$, mimicking the signature expected for an SMC-like dust law. 
We show that this provides a plausible mechanism for reconciling apparently contradictory results
in the literature and that, based on typical measurement
uncertainties, stellar mass provides a cleaner prediction of UV
attenuation than $\beta$. Although the situation at lower
stellar masses remains uncertain, we conclude that for  $2<z<3$ star-forming galaxies with
$\log(M_{\ast}/\Msun) \geq 9.75$, both the IRX$-\beta$ and
IRX$-M_{\ast}$ relations are well described by a Calzetti-like attenuation law.
\end{abstract}

\begin{keywords}
galaxies: high-redshift -- galaxies: evolution -- galaxies: starburst
-- galaxies: star formation -- submillimetre: galaxies
\end{keywords}

\section{Introduction}
Obtaining an accurate measurement of the evolution of the cosmic star-formation rate density remains a key goal in observational cosmology. 
A thorough, pre-ALMA, review of our understanding of the evolution of
the cosmic star-formation rate density (SFRD) from both the UV and far-infrared (FIR) was provided by \cite{mad14}. 
The data assembled for this review clearly demonstrate the dominant
contribution made by obscured star formation over the redshift range
$0<z<3$, and the remarkably good agreement between FIR and dust-corrected UV star-formation rate densities over
the same redshift interval. However, it is worth noting that the exact location of the peak in cosmic SFRD is still uncertain, as is the
steepness of the decline in SFRD to higher redshift
(e.g. \citealt{d17}; \citealt{novak17}). Moreover, due to the limited sensitivity of FIR
and radio observations, our current knowledge of the SFRD at $z\geq3$
is still largely confined to dust-corrected measurements of the integrated UV luminosity density derived from
studies of Lyman-break galaxies (e.g. \citealt{mclure13}; \citealt{derek16}; \citealt{fink16}; \citealt{stark16}).

Within this context, the vastly improved sensitivity at sub-mm and
mm-wavelengths provided by ALMA affords an unprecedented opportunity to
measure the contribution of obscured star formation at $z \geq 3$. 
For example, the individual detections and stacking results derived
from the deep 1.3-mm ALMA mosaic of the HUDF presented by \cite{d17} confirmed the dominant contribution of
obscured star formation (by a factor of 2--5) around the peak in cosmic SFRD at $z\simeq 2$. 
However, the results presented in \cite{d17} also indicated that a
sharp fall-off in obscured star formation at higher redshifts means that the SFRD is
dominated by the unobscured UV component at $z\geq 4$, a result subsequently confirmed by a stacking analysis of 
the deepest available sub-mm imaging from the SCUBA-2 Cosmology Legacy Survey \citep{nathan17}.

Despite the improved sensitivity provided by ALMA, it is clear that assembling an accurate picture of
the evolution of cosmic SFRD still requires a method for
reliably dust correcting rest-frame UV measurements. Traditionally this dust correction is made using the so-called Meurer relation:
$A_{1600~}=~1.99\beta~+~4.43$, which relates the observed UV spectral
slope (i.e. $\beta$ where $f_{\lambda}\propto \lambda^{\beta}$) of star-forming galaxies to the absolute level of
dust attenuation at 1600\AA\, \citep{m99}. 
The Meurer relation is itself derived from the IRX$-\beta$ relation, where
IRX (infrared excess) is usually defined as the ratio of infrared to
UV luminosity ($L_{\rm IR}/L_{\rm UV}$). Under the assumption that
star-forming galaxies all have similar {\it intrinsic} UV slopes, and
that their infrared luminosity arises from dust heated by the same UV continuum, it is possible to show
that IRX maps uniquely to $A_{1600}$ \citep{m99}. 

The great utility of the Meurer relation is that it allows
dust-corrected star-formation rates to be derived based on nothing
more than an apparently straightforward measurement of UV luminosity
and spectral slope and, as a consequence, its functional form and potential
evolution with redshift has been the focus of sustained interest
(e.g. \citealt{kong04}; \citealt{red10}; \citealt{over11};
\citealt{take12}; \citealt{lee12}; \citealt{capak15};
\citealt{salmon16}; \citealt{popping17}; \citealt{des17}).

However, two sets of observational results in the recent literature
have led to renewed interest in the validity and evolution of the IRX$-\beta$ relation.
Firstly, spectroscopic observations have revealed that the physical
conditions (including ionization parameter and metallicity) within star-forming galaxies at $z\geq 2$ are systematically different to
those of their low-redshift counterparts (e.g. \citealt{shap15};
\citealt{steidel16}; \citealt{ferg16}; \citealt{strom17}).
Secondly, several recent ALMA measurements of dust continuum emission in
$z\geq 2$ star-forming galaxies appear to suggest that at least some
high-redshift galaxies produce significantly less IR luminosity than
predicted by the Meurer relation (e.g. \citealt{capak15};
\citealt{bow16}; \citealt{majec16}; \citealt{pope17}; \citealt{fuda17}), and appear to follow an IRX$-\beta$ relation more
compatible with an SMC-like {\it extinction} law. Indeed, the recent results
presented by \cite{Reddy17} suggest that the bluest galaxies
at $1.5<z<2.5$ agree with the IRX$-\beta$ relation predicted by the
SMC extinction law, provided that their intrinsic UV slopes are
ultra-blue (i.e. $\beta_{\rm int}\simeq -2.6$).

The apparently low IRX values displayed by some high-redshift galaxies can be viewed as surprising from two
different perspectives. Firstly, theoretical studies (e.g. \citealt{witt2000};
\citealt{seon16}) have shown that dust properties that produce steep 
extinction laws (i.e. SMC-like) produce attenuation curves, appropriate for dust
correcting integrated galaxy photometry, which are much greyer in the
UV \footnote{Throughout the paper we refer to {\it
    extinction} as the absorption and scattering of light out of the line of
  sight, by dust between the object and the observer - as measured when
  deriving extinction laws in the MW and SMC. We refer to {\it
    attenuation} as the absorption and scattering of light out
  of/into the line of sight, appropriate for a galaxy where the stars and dust are mixed together. It is the scattering of light back into
  the line of sight which results in the attenuation law being greyer than the corresponding extinction law.}. 
Secondly, recent studies investigating the functional form of dust
reddening in high-redshift galaxies (e.g. \citealt{red15}; \citealt{scov15}; \citealt{ferg17}) have tended to conclude that dust attenuation
at $\lambda_{\rm rest}<2500\AA$ is similarly grey to the \cite{calz00}
starburst attenuation law, which predicts an IRX$-\beta$ relation 
virtually indistinguishable from the Meurer relation.

Adding to the confusing picture is the fact that 
several studies have produced individual or stacked sub-mm/mm
detections for $z\geq 3$ galaxies that are perfectly consistent with
the Meurer relation (e.g. \citealt{watson15}; \citealt{coppin15};
\citealt{d17}; \citealt{laporte17}). Motivated by these apparently contradictory results, in this paper we
investigate the IRX$-\beta$ and IRX$-M_{\ast}$ relations at $2\leq z
\leq 3$ using the deep ALMA 1.3-mm imaging of the HUDF presented by
\cite{d17}. The structure of the paper is as follows. In Section 2 we
describe the relevant data sets, our spectral energy distribution (SED) fitting and the
selection of our star-forming galaxy samples.
In Section~3 we briefly review the derivation of the IRX$-\beta$ relation and justify the
assumptions that we adopt throughout the rest of the analysis. 
In Section~4 we present our IRX$-\beta$ results at $2\leq z \leq 3$ 
and compare to the recent literature. In Section 5 we use a large, mass complete, sample of star-forming
galaxies at $2\leq z \leq 3$ to investigate the relationship between
UV spectral slope, UV dust attenuation and stellar mass.
In Section 6  we proceed to study the form of the IRX$-M_{\ast}$ relation at $2\leq z \leq 3$ in the HUDF, comparing our
results to recent literature studies. In Section 7 we investigate
possible biases in previous determinations of the IRX$-\beta$ relation, and present a
plausible mechanism for explaining apparent discrepancies in the
form of the IRX$-\beta$ and IRX$-M_{\ast}$ relations previously
presented. Finally, in Section 8 we provide a summary of our results and conclusions. Throughout the paper
magnitudes are quoted in the AB system \citep{oke83} and we
assume the following cosmology: $\Omega_{M}=0.3$, $\Omega_{\Lambda}=0.7$ and $H_{0}=70$ km s$^{-1}$ Mpc$^{-1}$.

\section{Data and SED fitting analysis}
In this section we provide a brief overview of the imaging data sets
employed in the current study and describe the SED fitting used to derive the key parameters required for the subsequent analysis.
In particular, we provide a full description of the methods used to fit the UV spectral slopes and define our
final samples of star-forming galaxies.

\subsection {ALMA data}
We use the ALMA 1.3-mm continuum image of the HUDF presented by \cite{d17}. The final mosaic was constructed from 45 separate
ALMA pointings and covers an area of $4.5$ arcmin$^{2}$,
overlapping with the ultra-deep ($m_{\rm AB}\simeq 30; 5\sigma)$,
near-IR {\it HST} imaging of the field (\citealt{ellis13}; \citealt{ill13}) obtained with Wide Field Camera 3 (WFC3/IR). 
The 1.3-mm mosaic has a uniform depth of 35 $\mu$Jy beam$^{-1} (1\sigma)$ at a spatial resolution of
$0.7$ arcsec. Full details of the observations and data reduction are provided in \cite{d17}, along with the multi-wavelength
photometry and derived physical properties for the 16 sources detected
with $S_{1.3}>120$ $\mu$Jy.

\subsection{HUDF optical/IR data}
The HUDF has been the focus of many ultra-deep imaging campaigns over the last decade, following-on from the
original Advanced Camera for Surveys (ACS) imaging obtained in 2003
(\citealt{beck06}). Of key importance to this study are the ultra-deep near-IR
observations obtained with WFC3/IR during the UDF09 \citep{bow11} and
UDF12 (\citealt{ellis13}; \citealt{koke13}) imaging campaigns.

The optical/near-IR imaging supplied by {\it HST}\, is supplemented by
ultra-deep ground-based UV and $K-$band imaging obtained with the
VIMOS \citep{non09} and HAWK-I \citep{font14} cameras on the VLT. 
Finally, the ultra-deep {\it Spitzer} IRAC imaging of
the field obtained via a series of different observing campaigns
(see \citealt{labbe15} for a summary) plays a crucial role in constraining the stellar masses of high-redshift galaxies.

In this study we make use of the GOODS-S photometry catalogue provided by the CANDELS survey team
\citep{guo13}. This catalogue provides PSF-homogenised 
photometry of the ACS and WFC3/IR imaging within the HUDF, in addition to resolution-matched photometry of
the lower spatial resolution ground-based and {\it Spitzer} IRAC
imaging derived using the {\sc tfit} deconfusion code \citep{laidler07}.
 
\subsection{CANDELS optical/IR data}
In order to expand the dynamic range of our star-forming galaxy sample we
supplement the HUDF data with the photometry for the wider GOODS-S field provided by \cite{guo13}. 
In addition, we also include the photometry assembled for the CANDELS UDS field by \cite{gal13}. 
The \cite{gal13} catalogue provides equivalent PSF-homogenised
UV-to-midIR photometry in the CANDELS UDS field to that provided by
\cite{guo13} for the CANDELS GOODS-S field. We refer the reader to \cite{gal13} for a full
description of the ground-based and space-based imaging incorporated within the catalogue.

\subsection{Ultra-VISTA optical/IR data}
Although constituting some of the best multi-wavelength data
available, the combined area of the CANDELS GOODS-S and UDS fields is $<0.1$ deg$^{2}$. 
Consequently, to significantly improve the
statistics of our star-forming galaxy sample at high stellar-masses
($10 \leq \log(M_{\ast}/\Msun) \leq 11.5$), we also incorporate
the photometry catalogue for the Ultra-VISTA (UVISTA) survey DR3 produced
for the $K-$band luminosity function study of \cite{mort17}. 
This catalogue includes $u^*g^{\prime}r^{\prime}i^{\prime}z^{\prime}$ photometry from the
T0007 data release of the CFHT Legacy Survey \citep{cfhtls}, deep
$z-$band photometry from Subaru Suprime-Cam \citep{furusawa16}, $YJHK$
imaging from UVISTA DR3 and {\it Spitzer} IRAC imaging from the
{\it Spitzer} Extended Deep Survey (SEDS; \citealt{ash13}) and {\it
  Spitzer} Large Area Survey (SPLASH; \citealt{splash}). The UVISTA 
photometry catalogue is split into components from the UVISTA deep and
wide strips. The deep strip component has a $5\sigma$ depth of
$K=24.8$ in a 2-arcsec diameter aperture. The wide
component has a $5\sigma$ depth of $K=23.9$.
Full details regarding the construction of the photometry catalogue can be found in \cite{mort17}.

\subsection{Photometric redshifts}
For the HUDF, GOODS-S and UDS data sets we
adopt the publicly available photometric redshifts derived by the
CANDELS team based on the \cite{guo13} and \cite{gal13} photometry
catalogues \citep{sant15}. These photometric redshifts are derived by combining the
results of many different photometric redshift analyses, using a
variety of different codes; see \cite{dahlen13} for full details.
For the UVISTA data set we adopt the photometric redshifts
previously derived by \cite{mort17}. These photometric redshifts are the median of five different photometric redshifts
runs, using three different template-fitting codes. As discussed in
detail by \cite{mort17}, the photometric redshifts for all four 
data sets are of equally high quality, with a typical value of
$\sigma_{\rm dz} = 0.02$ and a catastrophic outlier rate of 1--2 per cent (i.e. $|\rm{d}z| >0.15$).

\subsection{SED fitting analysis}
All of the UV-to-midIR photometry assembled from the HUDF, CANDELS and 
UVISTA data sets was analysed with the SED fitting code originally
developed by \cite{mclure11} and then subsequently updated as
described by \cite{derek15}. This code
fits stellar-population templates to the available photometry in
wavelength$-$flux density space to allow for the accurate handling of
photometric errors and non-detections. Provided with a set of
synthetic stellar-population models, the code derives the best-fitting
values of photometric redshift, stellar mass, star-formation rate and 
dust reddening. In addition, based on the best-fitting SED template 
the code also calculates rest-frame photometry in a wide variety of
different filters.

Throughout the main analysis pursued in this paper, the code used SED templates derived from
\cite{bc03} stellar population models, based on a
\cite{chab03} IMF, with dust reddening described by the \cite{calz00}
starburst attenuation law (with $A_{\rm V}$ in the range $0 \leq A_{\rm V} \leq 4.0$) and IGM absorption 
accounted for using the \cite{mad95} prescription. The fitting considered
standard $\tau-$model star-formation histories with values in the range $0.3$ Gyr $\leq \tau \leq 10.0$ Gyr, and metallicities of
either solar ($Z_{\odot}$) or one-fifth solar ($0.2Z_{\odot}$). During
the SED fitting process, templates were considered with ages between
50 Myr and the age of the Universe at the appropriate redshift.

The two physical parameters that are derived from the SED fitting and
used throughout the subsequent analysis are the stellar mass and
intrinsic UV spectral slope ($\beta_{\rm int}$). In order to check that our estimates are not biased by our choice of SED template set,
we experimented with a wide range of star-formation histories, including exponentially decaying, constant and exponential increasing.
These experiments demonstrated that the derived stellar masses are stable at the $\pm 0.1$ dex level and the values of $\beta_{\rm int}$
are stable at the $\pm 0.1$ level (using either Calzetti or SMC dust reddening).

As discussed in Section 7, the code was also used to explore the impact of low stellar-metallicity, nebular emission and stellar binaries by re-fitting the
HUDF star-forming galaxy sample using the `Binary
Population and Spectral Synthesis' (BPASSv2) stellar population models
(\citealt{jj16}; \citealt{stan16}). Once again, the derived values of
stellar mass and $\beta_{\rm int}$ were found to be stable to SED
template choice.

\subsection{Measuring the UV spectral slope}
\label{powerlaw}
Several different techniques have been advocated in the literature for 
measuring the UV spectral slope of star-forming galaxies based on
photometry alone; see \cite{sandy13} for a review. The techniques commonly
employed range from straightforward single-colour measurements
(e.g. \citealt{d13}; \citealt{bow12}), to applying the same UV windows
used to measure $\beta$ from UV spectra \citep{calz94} to the SED
templates that provide the best-fit to the available photometry \citep{fink12}. 

Here we adopt the technique employed by \cite{sandy14} for their study
of the colour-magnitude relation ($\beta-M_{\rm UV}$) at $z\simeq 5$. The best-fitting value of $\beta$ for each galaxy is determined by fitting the
photometry covering the wavelength range 1268\AA~$\leq~\lambda_{\rm
  rest}~\leq~2580$\AA\, with a series of pure power-law SED models, in
which IGM absorption is accounted for using the \cite{mad95}
prescription. The photometry for each galaxy was fit with a set of power-law models covering the range $-4.0 \leq \beta \leq
+4.0$ in steps of $\Delta \beta=0.01$, and the asymmetric
$\beta$ uncertainties were calculated from the $\Delta \chi^{2}=1$
interval around the best-fitting value.

\subsection{$\mathbf {UVJ}$ selection}
An essential requirement for the analysis presented in this paper is a
clean sample of $2 \leq z \leq 3$ star-forming galaxies free, as far
as possible, from potentially quiescent galaxies. To achieve this, we have adopted so-called $UVJ$ selection, which separates
quiescent and star-forming galaxies based on their positions on the
rest-frame $U-V$ vs $V-J$ colour-colour diagram. This method was first formalised by \cite{will09} and was shown to be
applicable out to redshifts of at least $z \simeq 2.5$ by \cite{whit11}. In addition to its ability to efficiently separate
star-forming and quiescent galaxies, an important advantage of $UVJ$
selection is that it is virtually independent of the choice of
stellar-population templates used when deriving the rest-frame $UVJ$ photometry.
This is notably not the case when separating star-forming and quiescent galaxies based
on a specific star-formation rate (sSFR) criterion.

Both \cite{will09} and \cite{whit11} agree on the best diagonal separation between the star-forming
and quiescent populations, but adopt slightly different additional
criteria to prevent contamination of the passive galaxies by both unobscured and dusty star-forming galaxies.
In order to minimize the contamination of our star-forming galaxy sample by potentially quiescent galaxies, we have adopted a 
hybrid of the \cite{will09} and \cite{whit11} criteria. Specifically,
any galaxy which meets all of the following criteria is classified as
{\it potentially} quiescent and therefore excluded from the
final samples:
\begin{equation}
\begin{split}
U-V& > 0.88(V-J)+0.59;\\
U-V& > 1.2;\\
V-J &  <  1.6.
\end{split}
\end{equation}
It should be noted that a straightforward adoption of either the \cite{will09} or \cite{whit11} criteria does not
significantly change the conclusions drawn from the star-forming
galaxy samples in the subsequent analysis.

\subsection{Star-forming galaxy samples at $\mathbf{2<z<3}$}
Throughout the rest of the analysis we will employ two samples of
$UVJ-$selected star-forming galaxies within the redshift interval $2 \leq z \leq 3$.
The first sample consists of galaxies selected from the area of the HUDF covered by
the ALMA 1.3-mm mosaic. This HUDF sample is mass complete down to a limit of
$\log(M_{\ast}/\Msun)~\geq~8.50$ and is used in Sections 4$-$6 for the  analysis of the IRX$-\beta$ and IRX$-M_{\ast}$ relations.

For the analysis of the relationship between UV spectral slope and
stellar mass, we enhance the dynamic range of the HUDF sample by
including $UVJ-$selected galaxies from the UDS, GOODS-S and UVISTA
data sets discussed above. To ensure mass-completeness, the GOODS-S and UDS samples are restricted to $\log(M_{\ast}/\Msun)\geq
8.95$ and $9.20$, respectively. For the UVISTA data set we impose
mass-completeness limits of $\log(M_{\ast}/\Msun)\geq 10.00$ and $10.35$ for the UVISTA-deep
and UVISTA-wide components, respectively. 

Following the application of the $UVJ$ selection criteria, a small number
of galaxies ($<1\%$) were identified that satisfied the $UVJ$
criteria as star-forming galaxies, but that have sSFRs which strongly
suggest that they are in fact quiescent (i.e. sSFR $\leq 0.1$~Gyr$^{-1}$). 
The vast majority of these apparent interlopers are located in an area of parameter space
($U-V>2.0; V-J>1.6$) where the $UVJ$ selection technique is prone to confusion
between dusty star-forming galaxies and genuinely passive
systems. After their removal the full, mass-complete, sample consists of 8273 star-forming galaxies at $2 \leq z \leq 3$,
spanning an area of $\simeq1$ deg$^{2}$ and the stellar-mass range: $8.50 \leq \log(M_{\ast}/\Msun)\leq 11.50$.

Finally, it is worth noting that the median sSFR for the full sample of $2 \leq z \leq 3$ star-forming
galaxies is $3.0~\pm~0.1$~Gyr$^{-1}$. This can be compared to the median
value of 3.2~Gyr$^{-1}$ predicted by the \cite{speagle14} meta-study of star-formation `main sequence' results. This comparison 
provides an additional level of confidence that the adopted $UVJ$ 
selection technique has successfully isolated a sample of galaxies consistent with populating the main sequence at $2 \leq z \leq 3$.

\section{The Infrared Excess}
In their original study of a sample of local starburst galaxies,
\cite{m99} defined the infrared excess IRX as:
\begin{equation}
\irx  \equiv \frac{F_{\rm FIR}}{F_{1600}},
\end{equation}
where $F_{1600}$ is the observed flux at $\lambda_{\rm rest}=1600\AA$,
defined as $F_{1600}=\nu S_{\nu}$, and the far infrared flux is defined as:
\begin{equation}
{\rm F_{FIR}}=1.25\left[ F(60)+F(100)\right],
\end{equation}
where $F(60)$ and $F(100)$ are the fluxes measured by {\it IRAS} in the
60 $\mu$m and 100 $\mu$m bands \citep{hel88}. Under the assumption that the infrared luminosity of a galaxy represents reprocessed emission
from dust heated by the UV photons produced by an actively star-forming stellar population, \cite{m99} showed
that IRX can be accurately approximated as:
\begin{equation}
\irx = B\left(10^{0.4A_{1600}}-1\right),
\end{equation}
\noindent
where $A_{1600}$ is the attenuation at $\lambda_{\rm rest}=1600\AA$ in
magnitudes, and the multiplicative
factor was calculated by \cite{m99} to have a value of $B=1.19\pm{0.20}$.
The multiplicative factor is actually the ratio of two bolometric correction factors:
\begin{equation}
B = \frac{\rm BC(1600)}{\rm BC(FIR)},
\end{equation}
where the numerator effectively converts between $F_{1600}$ and total
flux available to heat the dust, and the denominator converts between $F_{\rm FIR}$ and the bolometric IR flux emitted by the dust.

Throughout the analysis presented in this paper, as is common practice
for studies of high-redshift galaxies, we adopt the following IRX definition:
\begin{equation}
\irx  \equiv \frac{L_{\rm IR}}{ L_{\rm UV}},
\end{equation}
where $L_{\rm IR}$ is the total IR luminosity ($8$--$1000\mu$m) and $L_{\rm UV}$ is $\lambda L_{\lambda}$
evaluated at 1600\AA. Using this definition, ${\rm BC(FIR)}=1$ and the multiplicative factor is simply $B=\rm BC(1600)$. 
Here we take an empirical approach to setting the value of $\rm
BC(1600)$, adopting the median value of $\rm BC(1600)=1.71\pm{0.05}$ derived from the
SED fits to the full $2\leq z \leq 3$ sample of star-forming galaxies
defined in Section 2. Consequently, throughout this study we  adopt
the following relationship between IRX and $A_{1600}$:
\begin{equation}
\irx = 1.71\left(10^{0.4A_{1600}}-1\right),
\label{irx}
\end{equation}
\noindent
noting that this normalization is consistent with that adopted by many
previous studies, which typically fall within the range
$1.66<{\rm BC(1600)} <2.08$ (e.g. \citealt{m99}; \citealt{red10};
\citealt{over11}; \citealt{ferg17}). 

In practice, when deriving IRX values for both individual objects and
stacked data, we calculate the ratio of SFR$_{\rm IR}$ to SFR$_{\rm UV}$.
To calculate SFR$_{\rm IR}$ we adopt the results of the template
fitting performed on the full long-wavelength (24 $\mu$m$-1.3$ mm)
SEDs of the detected ALMA objects presented in \cite{d17}. For the objects within the redshift interval $2<z<3$, the results of
this SED fitting indicate that
\begin{equation}
{\rm SFR}_{\rm IR}/\Msun {\rm yr}^{-1} \simeq 0.3\times
(S_{1.3}/\mu {\rm Jy}),
\end{equation}
where $S_{1.3}$ is the 1.3-mm ALMA flux density.
This calibration is virtually identical to first calculating $L_{\rm IR}$ from the 1.3-mm flux, assuming an optically thin
modified blackbody spectrum with a dust temperature of $T_{\rm dust}=35$ K and
a dust emissivity index of $\beta_{\rm dust}=1.6$, and then applying:
\begin{equation}
\log\left( {\rm SFR}_{\rm IR}/\Msun {\rm yr}^{-1}
\right)=\log\left( L_{\rm IR} /{\rm W}\right)-36.44,
\end{equation}
\noindent
which is the SFR$_{\rm IR}$ calibration from \cite{ken12}, converted to a \cite{chab03} IMF. 

To calculate the value of SFR$_{\rm UV}$ we adopt the following calibration:
\begin{equation}
\log\left({\rm SFR}_{\rm UV}/\Msun {\rm
    yr}^{-1}\right)=\log\left(L_{\rm UV} /{\rm W}\right)-36.44,
\end{equation}
where $L_{\rm UV}$ is calculated from the absolute magnitude at 1500\AA\,
derived from the best-fitting SED templates to the available UV-to-midIR photometry. This calibration is taken from
\cite{mad14}, and has again been converted to a \cite{chab03} IMF. It
can be seen from equations (9) and (10) that with our choice of SFR calibrations
\begin{equation}
\irx  \equiv \frac{L_{\rm IR}}{ L_{\rm UV}} \equiv \frac{{\rm
    SFR}_{\rm IR}}{{\rm SFR}_{\rm UV}}.
\end{equation}
\noindent
As a result, we will use IRX and the ratio of IR to UV
star-formation rate interchangeably throughout the rest of the paper.

\begin{figure*}
\includegraphics[width=16cm,angle=0]{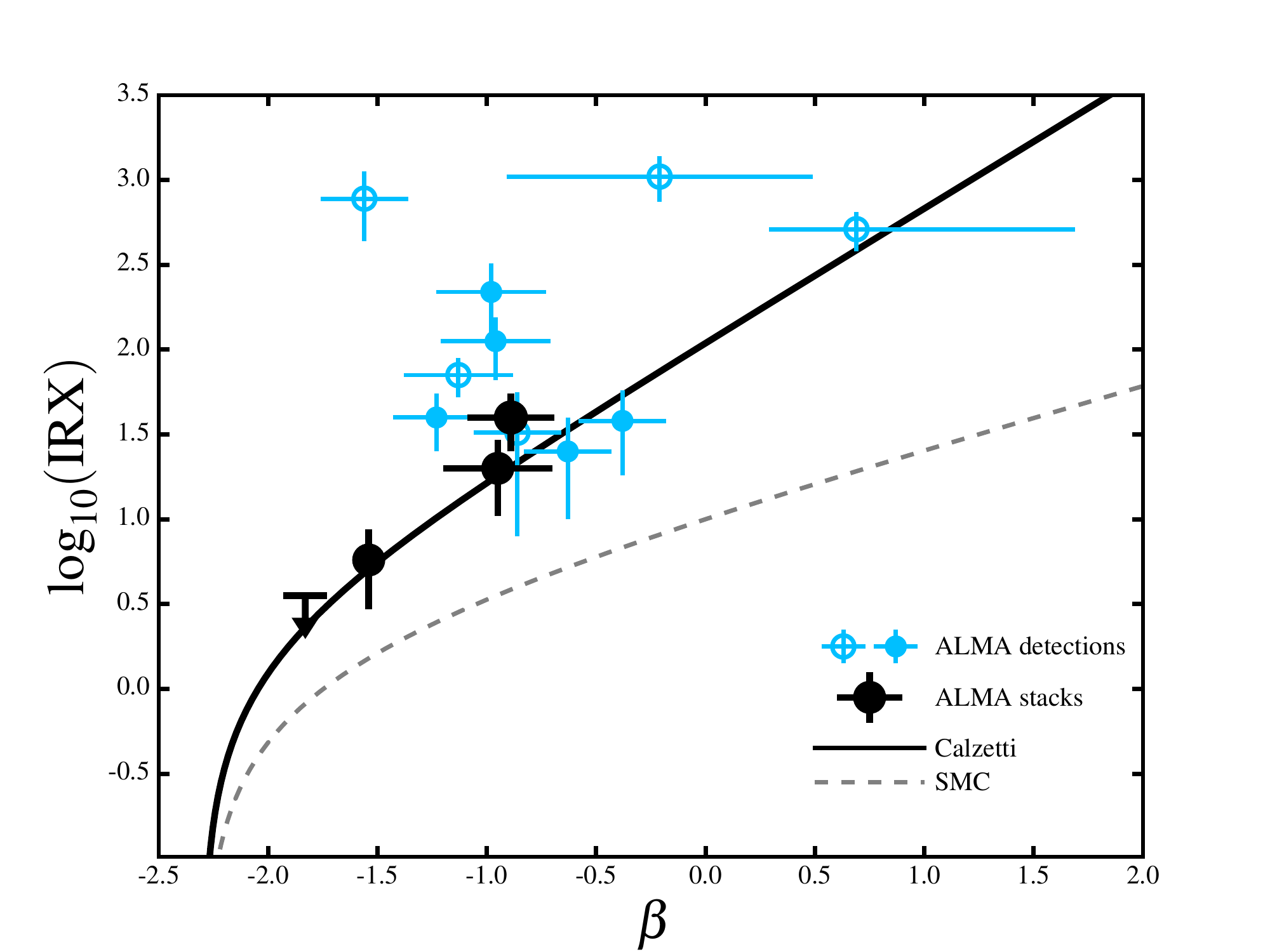}
\caption{The IRX$-\beta$ relation for $2 \leq z \leq 3$ star-forming galaxies
within the HUDF. The black upper limit ($2\sigma$) corresponds to a stack of 
$2 \leq z \leq 3$ galaxies from our HUDF star-forming galaxy sample with
stellar masses in the range $8.50\leq\log(M_{\ast}/\Msun)<9.25$. 
The black data-point at $\beta = -1.54\pm{0.05}$ corresponds to a
stack of $2 \leq z \leq 3$  HUDF star-forming galaxies in the mass
range $9.25\leq\log(M_{\ast}/\Msun)<10.00$. The two black data points at $\beta\simeq
-0.9$ both correspond to stacks within the mass range $10.00\leq\log(M_{\ast}/\Msun)<10.75$. The higher of the two data points is a
stack of all galaxies within the stellar mass bin, including sources
that were individually detected at $S_{1.3}>120$ $\mu$Jy. The lower of
the two data points is a stack of those objects that were not
individually detected. The solid black line is the predicted IRX$-\beta$ relation under the assumption that
the underlying dust curve follows the \protect\cite{calz00}
attenuation law. The dashed grey line is the predicted IRX$-\beta$ relation under the
assumption that the underlying dust curve follows the SMC extinction
law (e.g. \protect\citealt{gord03}). The blue data points are ten galaxies with redshifts in the range
$1.72<z<3.08$, which were identified in the 1.3-mm ALMA mosaic of the
HUDF with $S_{1.3}>120$ $\mu$Jy by \protect\cite{d17}. The solid blue
data points show the ALMA-detected objects which are included in the
black stacked data points (i.e. those objects with $2\leq z \leq 3$
and $8.50\leq\log(M_{\ast}/\Msun)<10.75$). The open blue data points show the
ALMA-detected objects which are not included in the stacked data
points due to having a redshift outside the range $2\leq z \leq 3$, or $\log(M_{\ast}/\Msun)>10.75$.}
\end{figure*}

\begin{figure}
\begin{center}
\includegraphics[width=5.8cm,angle=0]{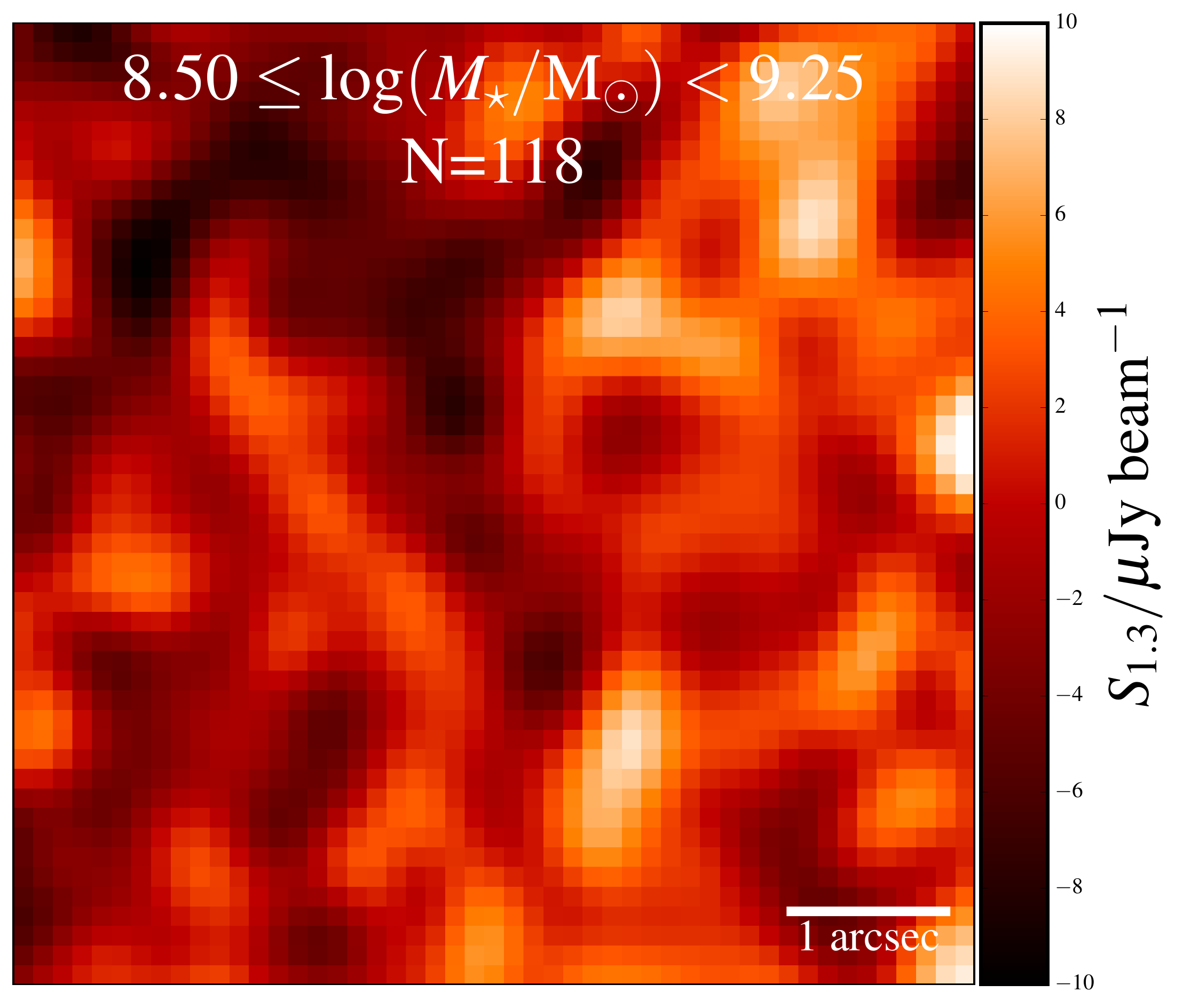}

\vspace{-0.12cm}
\includegraphics[width=5.8cm,angle=0]{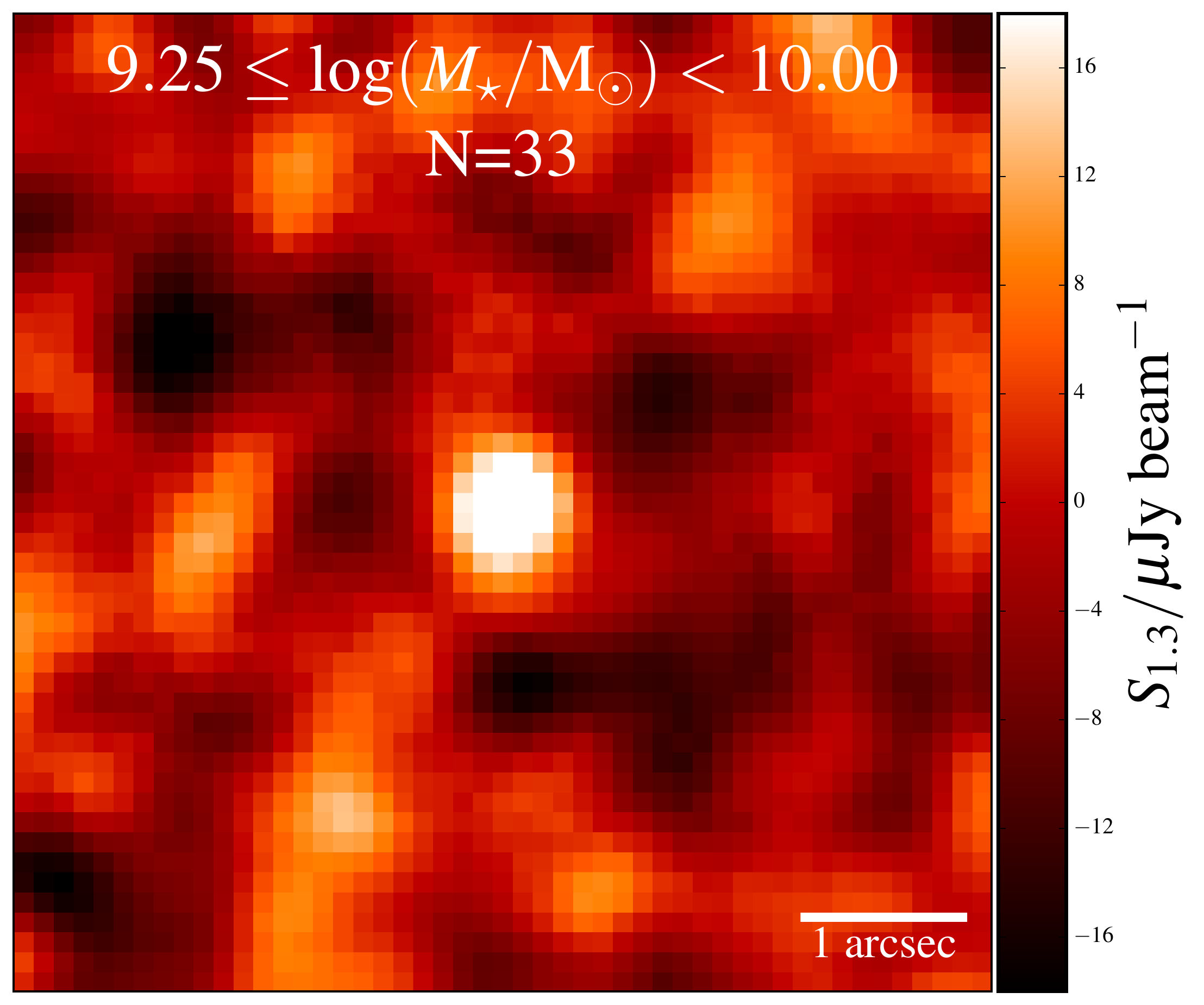}

\vspace{-0.10cm}
\includegraphics[width=5.8cm,angle=0]{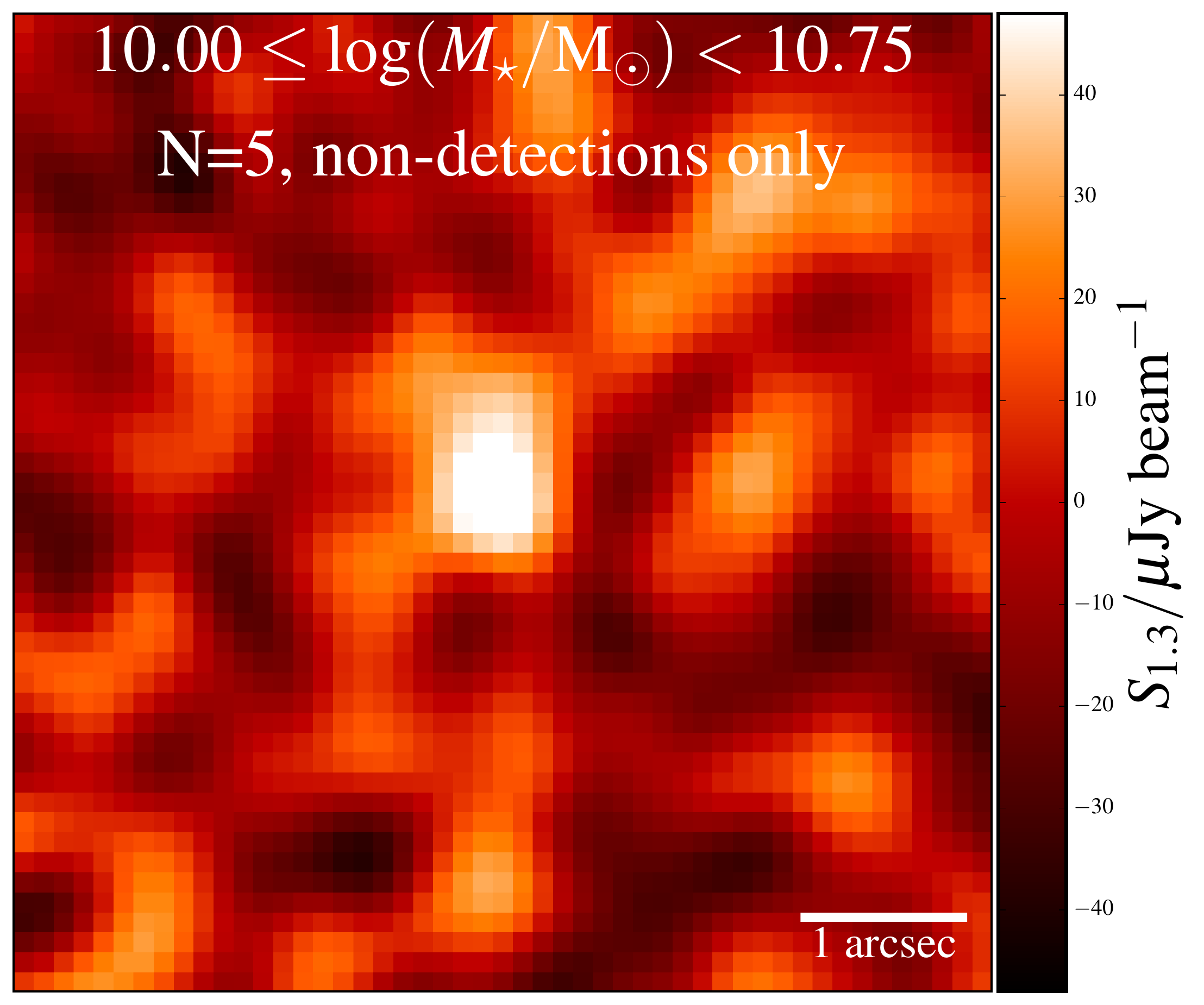}

\vspace{-0.09cm}
\includegraphics[width=5.8cm,angle=0]{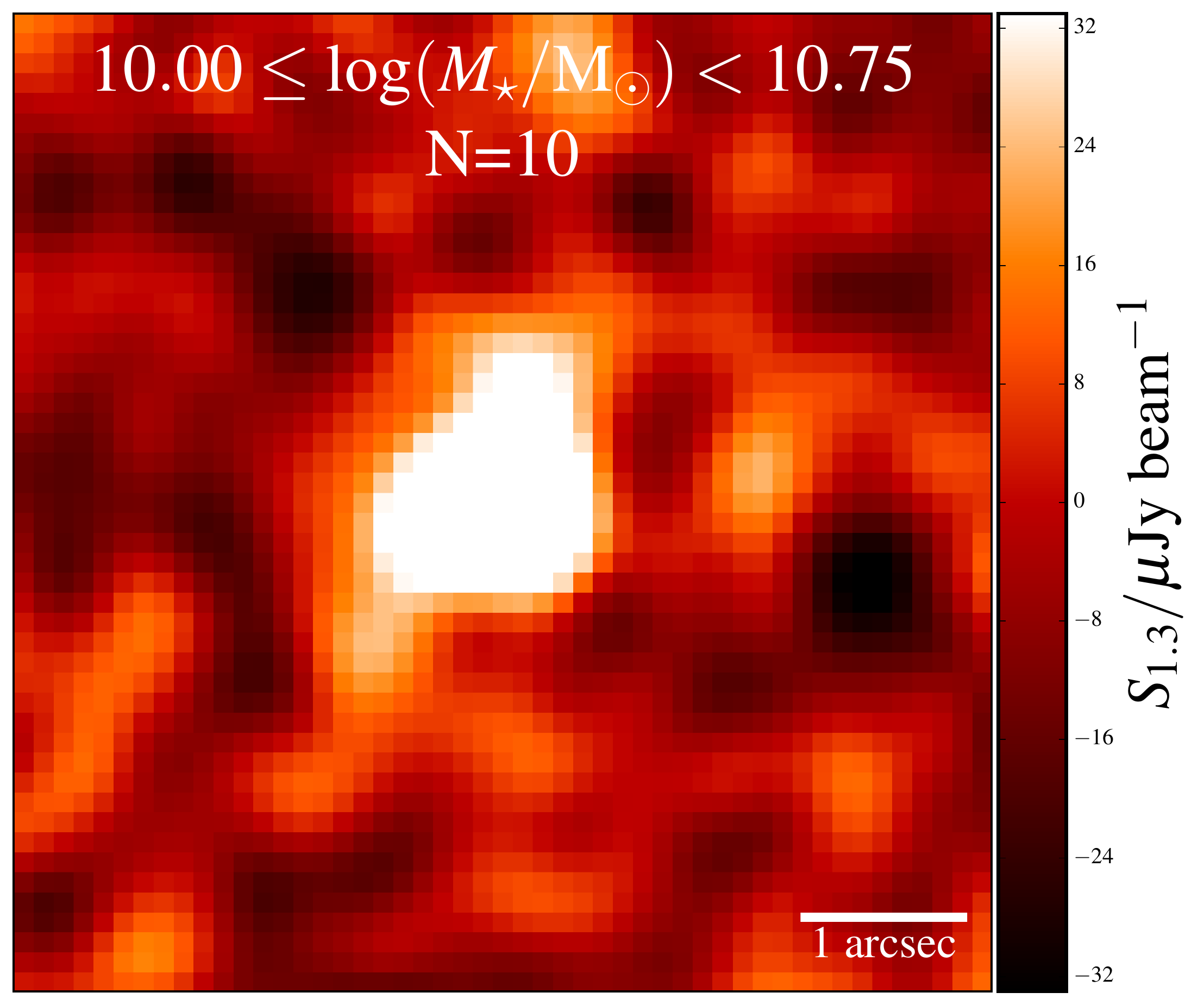}
\end{center}
\caption{Mean ALMA 1.3-mm stacks of $2.0<z<3.0$ star-forming galaxies
  in the HUDF. The top stack is a non-detection ($2\sigma$ upper limit
  $\simeq 6$~$\mu$Jy) for galaxies in the stellar-mass range $8.50 \leq  \log(M_{\ast}/\Msun)<9.25$.
The second stack is a $6\sigma$ detection for galaxies in the
stellar-mass range $9.25\leq\log(M_{\ast}/\Msun)<10.00$. The third stack is
  a $4\sigma$ detection for galaxies in the mass range $10.00\leq\log(M_{\ast}/\Msun)<10.75$, which are not 
  {\it individually} detected at $S_{1.3}>120$ $\mu$Jy ($3.5\sigma$ significance). The
  final stack consists of all galaxies within this last mass range,
  including 5/16 of the $S_{1.3}>120$ $\mu$Jy sources identified by \protect\cite{d17}.}
\end{figure}

\subsection{Connecting IRX and dust reddening}
In principle, equation (7) allows the value of IRX to be estimated based solely on a measurement
of $A_{1600}$. However, given that a direct measurement of $A_{1600}$ is impractical, in practice it is necessary to make
the further assumption that before dust attenuation, the intrinsic UV spectral slopes (i.e. $\beta_{\rm int}$) of all actively star-forming
galaxies are very similar. If this assumption is valid then, for a given assumed form of the underlying attenuation law, it is possible
to calculate $A_{1600}$ based on a measurement of the {\it observed} UV spectral slope (i.e. $\beta_{\rm obs}$), since:
\begin{equation}
A_{1600}=\frac {dA_{1600}}{d\beta}(\beta_{\rm obs}-\beta_{\rm int}),
\end{equation}
where the slope of the reddening law has the values
\begin{equation}
\frac{dA_{1600}}{d\beta} = 1.99,\quad\quad \frac{dA_{1600}}{d\beta} = 1.97, \quad\quad\frac{dA_{1600}}{d\beta} = 0.91,
\end{equation}
for the \cite{m99} relation, the \cite{calz00} starburst attenuation law and
the SMC extinction law, respectively. The slope of the SMC extinction 
law adopted here is based on the derivation of \cite{gord03},
although earlier studies of UV extinction in the SMC by
(e.g. \citealt{prev84}) produce a virtually identical value for the
slope. We note that our calculations of the dust law slopes assume that
$\beta$ is measured between $\lambda_{1}=1276$\AA\, and $\lambda_{2}=2490$\AA, the centres of the first and last UV windows
defined by \cite{calz94}. Other values quoted in the literature are the result of different choices of wavelength anchors. For example,
changing to $\lambda_{1}=1600$\AA\, results in dust law slopes of
$dA_{\mbox{\tiny 1600}}/d\beta \simeq 2.2$ and $dA_{\mbox{\tiny
    1600}}/d\beta \simeq 1.2$ for the \cite{calz00} attenuation and SMC extinction laws respectively.

It can be seen from equation (13) that the slope of the reddening law implied by the \cite{m99} relation and the
\cite{calz00} attenuation law are virtually identical. As a consequence, throughout this paper we will consistently plot IRX curves corresponding to
the \cite{calz00} attenuation law, noting that the IRX curves corresponding to the \cite{m99} relation are virtually indistinguishable.

As highlighted in the introduction,  dust properties that produce a steep {\it extinction} law (e.g. SMC or MW), will inevitably produce an
{\it attenuation} law that is much shallower\footnote{Where
  steep and shallow refer to large or small values of
  $\frac{d\beta}{dA_{\mbox{\tiny 1600}}}$.}
(i.e. greyer). Therefore, when dealing with integrated galaxy photometry, it is unlikely that
the effective attenuation curve will be as steep as the SMC extinction
law. However, despite this, throughout the paper we will also
consistently plot predicted IRX curves based on the SMC extinction law, for two reasons.
Firstly, as noted by \cite{pet98}, the SMC curve does serve as a useful lower-limit to the amount of UV attenuation.
Secondly, as noted in the introduction, there have been several claims
in the recent literature that the IRX$-\beta$ relation followed by high-redshift galaxies is closer to the predictions of the SMC extinction law, than
the original Meurer/Calzetti relation.

For completeness, our final adopted IRX relations are as follows:
\begin{equation}
\irx = 1.71\left(10^{0.4\times1.97\left(\beta_{\rm obs}-\beta_{\rm
        int}\right)}-1\right) \quad {\rm [Calzetti]}\\
\end{equation}
\begin{equation}
\irx = 1.71\left(10^{0.4\times0.91\left(\beta_{\rm obs}-\beta_{\rm int}\right)}-1\right) \quad {\rm [SMC]} 
\end{equation}
where $\beta_{\rm obs}$ is the observed value of the UV spectral slope and
$\beta_{\rm int}$ is the assumed intrinsic value. Throughout this analysis,
we adopt $\beta_{\rm int}=-2.3$, which is the median value of the
intrinsic UV spectral slope based on the SED fitting of our sample of 
$2<z<3$ star-forming galaxies in the HUDF. We note that this is
similar to the canonical value of $\beta_{\rm int}=-2.23$ adopted by \cite{m99}.

\begin{figure*}
\includegraphics[width=16cm]{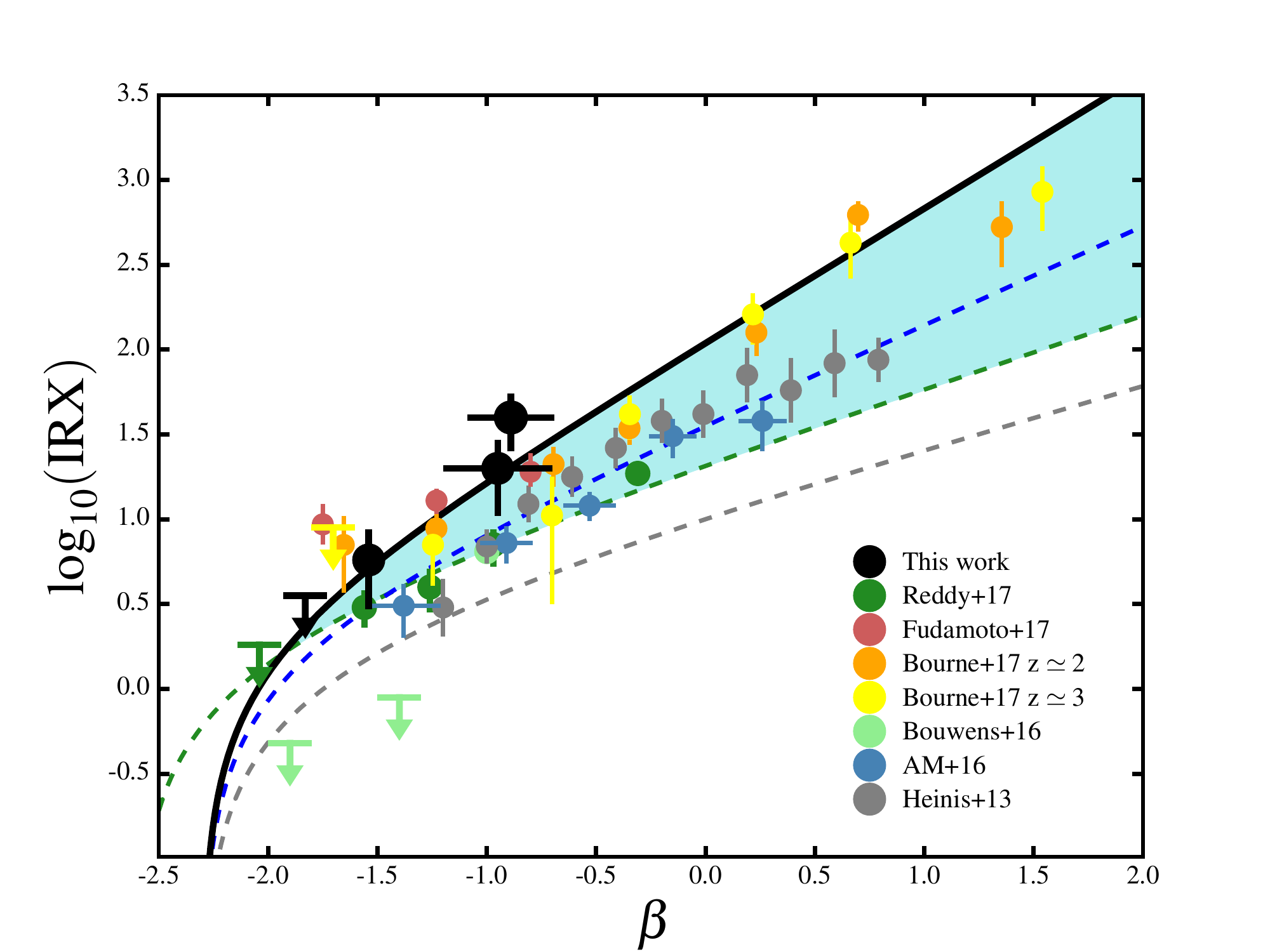}
\caption{A compilation of IRX$-\beta$ results from the recent
  literature for star-forming galaxies within the redshift interval
  $1.5\leq z \leq 3$. In addition to the new results
  derived in this work (black data points), we have plotted the
  results from the following studies: \protect\cite{heinis13}, \protect\cite{am16},
\protect\cite{bow16}, \protect\cite{nathan17}, \protect\cite{fuda17}
and \protect\cite{Reddy17}. As in Fig. 1, the solid black and dashed grey lines show
the expected IRX-$\beta$ relations for Calzetti-like and SMC-like dust
curves. The dark green dashed line shows the IRX$-\beta$ relation
derived by \protect\cite{Reddy17}. The blue dashed line shows the
IRX$-\beta$ relation for a dust curve with a slope in the UV that is
intermediate to the Calzetti and SMC-like relations (see text for
details). The blue shading highlights the region into which observational
measurements of the IRX$-\beta$ relation are likely to scatter, even
when the underlying IRX$-\beta$ relation is close to the Calzetti-like
prediction (black line); see Section 7 for a full discussion.}
\end{figure*}

\section{IRX versus UV slope}
In Fig.~1 we show the IRX$-\beta$ relation for
star-forming galaxies at $z\simeq 2.5$, based on our analysis of the
deep ALMA 1.3-mm mosaic of the HUDF.
The blue data points show the ten galaxies within the redshift range
$1.72<z<3.08$ that were detected in the ALMA mosaic with $S_{1.3}\geq
120$ $\mu$Jy. The IRX values for these objects are based on the SFR$_{\rm
  IR}$ and SFR$_{\rm UV}$ estimates derived by \cite{d17}, and are
plotted with $\beta$ values based on our power-law SED fitting.
The solid blue data points show the ALMA-detected objects which are included in the
black data points derived via stacking into the ALMA mosaic (see Section 4.1). The open blue data points show the
ALMA-detected objects which are not included in the stacked data
points due to having a redshift outside the range $2\leq z \leq 3$, or $\log(M_{\ast}/\Msun)>10.75$.
The solid black and dashed grey lines show the expected IRX$-\beta$
relations for Calzetti and SMC-like dust laws (equations 14 and 15)
under the assumption that $\beta_{\rm int}=-2.3$.

As can be seen from Fig.~1, those objects that were
individually detected in the HUDF 1.3-mm mosaic are either consistent with the Calzetti IRX$-\beta$
relation, or lie significantly above it. It is of course entirely expected that
heavily-obscured objects will lie above the IRX$-\beta$ relation, simply because the fundamental dust-screen assumption underlying the
derivation of equation (7) will not hold for the most heavily obscured
sources. Indeed, many of the individually detected objects in the ALMA
1.3-mm mosaic show spatial off-sets beween the obscured and unobscured star-formation \citep{whipu16}.

\subsection{Stacking results}
The black data points in Fig.~1 show the results of
stacking the ALMA data for those $2.0 \leq z \leq 3.0$ star-forming
galaxies in the HUDF with stellar masses in the following ranges
(left-to-right): $8.50 \leq \log(M_{\ast}/\Msun) < 9.25$, $9.25 \leq
\log(M_{\ast}/\Msun) < 10.00$ and $10.00 \leq \log(M_{\ast}/\Msun)
<10.75$. For each stellar-mass bin, we first calculate the individual
values of IRX for each object entering the bin (including negative
values of IRX corresponding to negative ALMA fluxes), and then plot the unweighted mean value of IRX at the corresponding unweighted mean value of $\beta$.

The final mass bin has been split into two data points, the
higher of which shows the result of stacking all ten galaxies within the  $10.00 \leq \log(M_{\ast}/\Msun) <10.75$ mass bin. Five of these ten
galaxies (solid blue data points in Fig.~1) were individually detected at $S_{1.3}\geq 120$ $\mu$Jy ($\geq
3.5\sigma$) by \cite{d17}. The lower data point in the figure corresponds to a stack of the five galaxies within this mass
bin that were not individually detected at $\geq 3.5\sigma$ by
\cite{d17}. The data point for the $8.50 \leq \log(M_{\ast}/\Msun) <
9.25$ stellar-mass bin is a robust $2\sigma$ upper limit
(i.e. measured value $+2\sigma$). The stacked images corresponding to the four black data points are shown in Fig.~2. 

As can be seen from Fig.~1, the first three
stellar-mass bins are all perfectly consistent with the IRX$-\beta$
relation expected for Calzetti-like dust attenuation, although the
upper limit is obviously also consistent with an SMC-like dust
curve. The stacked data point which includes five objects detected
with $S_{1.3}\geq 120$ $\mu$Jy (solid blue data points in Fig.~1), lies a factor of $\simeq 2$ above the
Calzetti IRX$-\beta$ relation. This is entirely consistent 
with the results presented in \cite{d17}, which indicated
that SED template fitting using the \cite{calz00} attenuation law was capable of recovering the total (i.e. UV+IR) star-formation rate for
galaxies with $\log(M_{\ast}/\Msun)~\leq~10.3$, but under-estimated
the total star-formation rate by approximately a factor of two at higher masses.

The results presented in Fig.~1 for both individual
objects and stellar-mass stacks strongly suggest that the IRX$-\beta$
relation for star-forming galaxies at $z\simeq 2.5$ is fully
consistent with that expected for Calzetti-like dust attenuation. In contrast, there is no indication from these results that the
effective dust attenuation at this epoch is as steep as the SMC extinction curve (at least for $\beta_{\rm int}\simeq -2.3$).
This surprisingly clear result is in apparent contrast to the wide variety of different results that have been presented in the recent literature.

\subsection{Comparison to recent literature results}
In Fig.~3 we show a compilation of results from a
number of different studies of the IRX$-\beta$ relation 
taken from the recent literature (\citealt{heinis13}; \citealt{am16};
\citealt{bow16}; \citealt{nathan17}; \citealt{fuda17}; \citealt{Reddy17}).
Crucially, all of these studies focused on samples of
star-forming galaxies within the redshift interval $1.5\leq z \leq 3$,
and featured infrared luminosities derived from
{\it Herschel}, SCUBA-2 or ALMA data.

In common with this study, the results of both \cite{bow16} and
\cite{fuda17} are based on mm-wavelength imaging with ALMA. In their study, \cite{bow16} investigated the IRX$-\beta$ relation of
$1.5<z<3$ Lyman break galaxies (LBGs), based on an independent 1.2-mm
ALMA mosaic within the HUDF. This mosaic is slightly deeper than the 1.3-mm mosaic studied
here, but covers an area which is approximately four times
smaller. The recent study of \cite{fuda17} is based on ALMA Band 6 continuum observations
of 67 massive star-forming galaxies in the COSMOS field with a median redshift of $3.2$. 

\cite{am16} studied the IRX$-\beta$ relation using a sample of LBGs at
$2.5<z<3.5$ within the COSMOS field, using stacked {\it Herschel} PACS
(100 and 160~$\mu$m), SPIRE (250, 350 and 500 $\mu$m) and AzTEC 1.1-mm
imaging to derive infrared luminosities. Similarly, \cite{heinis13} used stacked SPIRE data
(250, 350 and 500 $\mu$m) to study the IRX$-\beta$ relation for a sample of UV-selected galaxies at $z\simeq 1.5$ within the COSMOS field.

In their recent study, \cite{Reddy17} investigated the IRX$-\beta$ relation
for $UVJ-$selected star-forming galaxies at $1.5<z<2.5$, exploiting
UV imaging from the Hubble Deep UV Legacy survey within the two
GOODS fields. To derive their infrared luminosities, \cite{Reddy17} stacked
into the  available short-wavelength {\it Herschel} PACS data 
(100 and 160 $\mu$m).

Finally, the recent study of \cite{nathan17} stacked the deepest
available 450-- and 850--$\mu$m imaging from the SCUBA-2 Cosmology
Legacy Survey (CLS; \citealt{cls}), in combination with 100, 160 and 250--$\mu$m {\it Herschel}
imaging, to determine the infrared luminosities of $UVJ-$selected
star-forming galaxies within three of the CANDELS survey fields.
Using UV data from catalogues constructed by the 3D-HST survey team
(\citealt{brammer12}; \citealt{mom16}), \cite{nathan17} studied the
form of the IRX$-\beta$ relation for star-forming
galaxies over the redshift interval $0.5<z<6.0$. In Fig.~3 we plot the IRX-$\beta$ results for their galaxy
sub-samples at $z\simeq 2$ and $z\simeq 3$.

Taken at face value, the compilation of literature results shown in
Fig.~3 indicates that the IRX$-\beta$ relation
displayed by $1.5~<~z~<~3.0$ star-forming galaxies lies roughly equidistant from the expected relations for Calzetti-like and SMC-like dust laws.
Consequently, there are a limited number of options available to
account for differences in the observed data, provided that one is happy to accept
the basic tenet that the IRX$-\beta$ relation results from the dust
attenuation of star-forming galaxies, all of which have very similar {\it intrinsic} UV slopes.

The first option is to assume that the intrinsic UV slope of
star-forming galaxies at this epoch is $\beta_{\rm int}\simeq -2.3$, as
indicated by the SED fitting performed here. In this case, the most straightforward
interpretation of the data is that the effective attenuation law at
this epoch has a slope intermediate to that of the \cite{calz00} and
SMC dust laws (i.e. $dA_{\mbox{\tiny 1600}}/d\beta \simeq 1.45$).
As an illustration, the blue dashed curve in Fig.~3
shows the IRX$-\beta$ relation corresponding to equation (7), combined with
$A_{1600}=1.45\beta-3.34$. This IRX$-\beta$ relation corresponds to a dust law with slope
$dA_{\mbox{\tiny 1600}}/d\beta \simeq 1.45$ and $\beta_{\rm int}=-2.3$. It
can be seen that this provides a reasonable fit to the data from the
literature over a wide range of observed UV slope, and is similar to the
IRX$-\beta$ relations derived by \cite{heinis13} and \cite{am16}.

The second option is to adopt an SMC-like dust law and assume that the intrinsic UV-slopes of galaxies at this epoch are extremely blue. 
This is the conclusion reached by \cite{Reddy17}, based on their recent
study of star-forming galaxies at $1.5<z<3.0$. The green dashed line in Fig.~3 shows the preferred
solution of \cite{Reddy17}, which is based on SMC-like reddening
($\delta A_{\mbox{\tiny 1600}}/\delta\beta =1.07$) and an assumed value of $\beta_{\rm int}=-2.62$. 
\cite{Reddy17} argue that the relationship between IRX and $\beta$
effectively de-couples at UV slopes redder than $\beta_{\rm obs}\simeq
-0.5$, although it can be seen that their IRX-$\beta$ relation
systematically underestimates the observed IRX values from most other studies at $\beta > -1.0$.

One of the motivations given by \cite{Reddy17} for their adoption of
$\beta_{\rm int}=-2.62$, was that a BPASSv2 stellar population model with
this intrinsic value of UV slope is a good match to the stacked UV
spectra of 30 star-forming galaxies at $z\simeq 2.4$ presented by \cite{steidel16}. 
In Section 7 we investigate the impact of SED fitting with the BPASSv2 
stellar-population models, including the contribution from nebular
emission, and find that the typical values of $\beta_{\rm int}$ are significantly redder.

There is also a third option that is worth considering. It is 
possible that the true IRX$-\beta$ relation of galaxies at this epoch
does lie close to the Calzetti-like prediction, with $\beta_{\rm int}\simeq -2.3$.
In this scenario, the observed IRX$-\beta$ data from the literature is
explained by a systematic bias towards lower IRX values at red values of $\beta$.
As explored in more detail in Section 7, a bias must be
introduced at some level by the process of binning-up galaxy samples by
$\beta_{\rm obs}$ and, for plausible levels of uncertainty in individual $\beta_{\rm obs}$ measurements, might be expected 
to scatter data into the shaded region highlighted in Fig.~3. In this
scenario, our new ALMA-based IRX$-\beta$ results are less biased than
some previous determinations in the literature, simply because they are based
on binning in stellar mass rather than $\beta$, a conclusion supported
by the simulation work presented in Section 7.

Fortunately, it is possible to gain crucial information on which of
these three possibilities provides the best explanation of the
observed IRX$-\beta$ data, by studying the relationship
between stellar mass and UV attenuation in more detail.

\begin{figure*}
\includegraphics[trim=0.5cm 0.2cm 3.0cm 1.0cm, clip, width=0.49\textwidth]{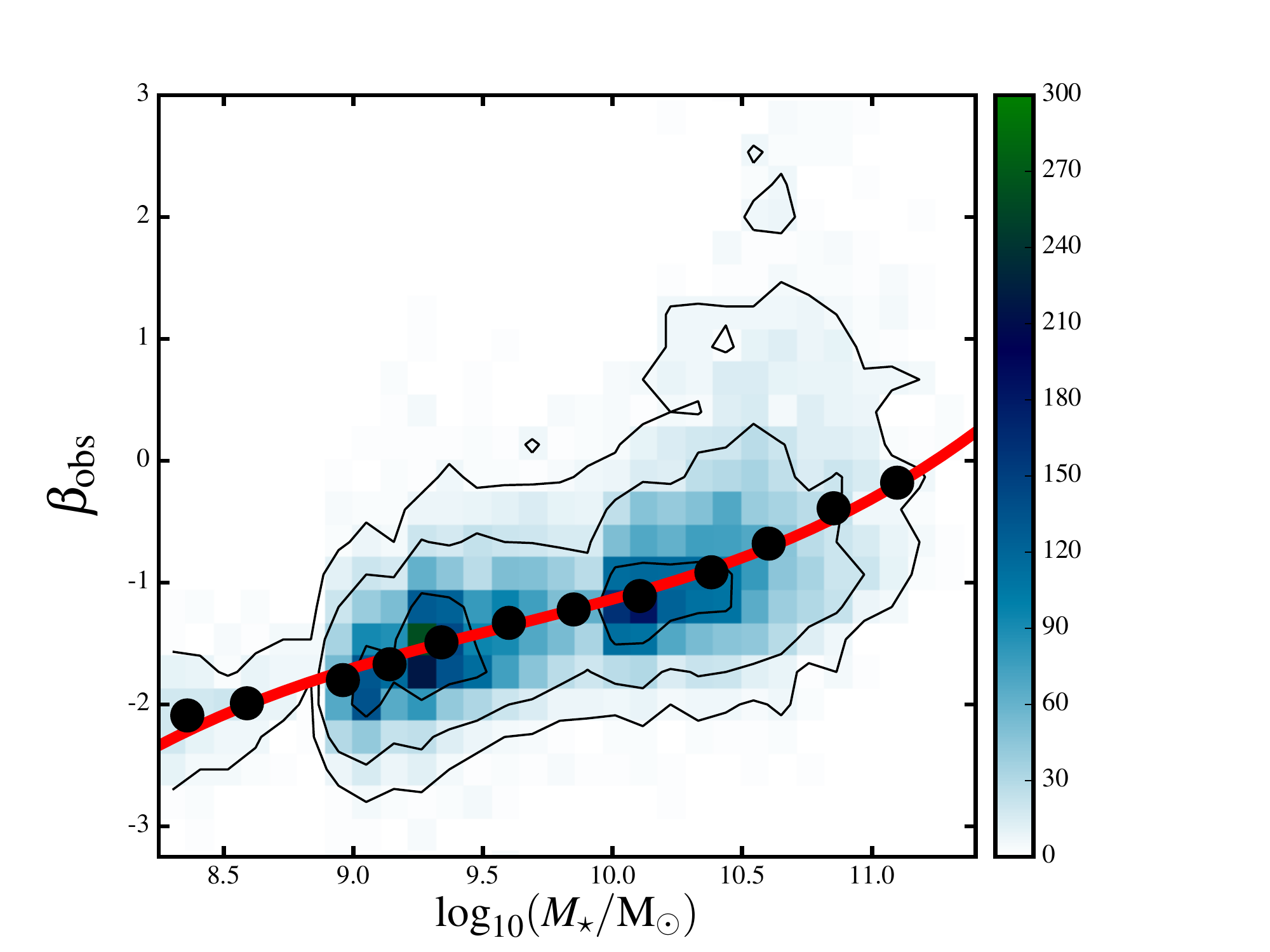}
\includegraphics[trim=0.5cm 0.2cm 3.0cm 1.0cm, clip, width=0.49\textwidth]{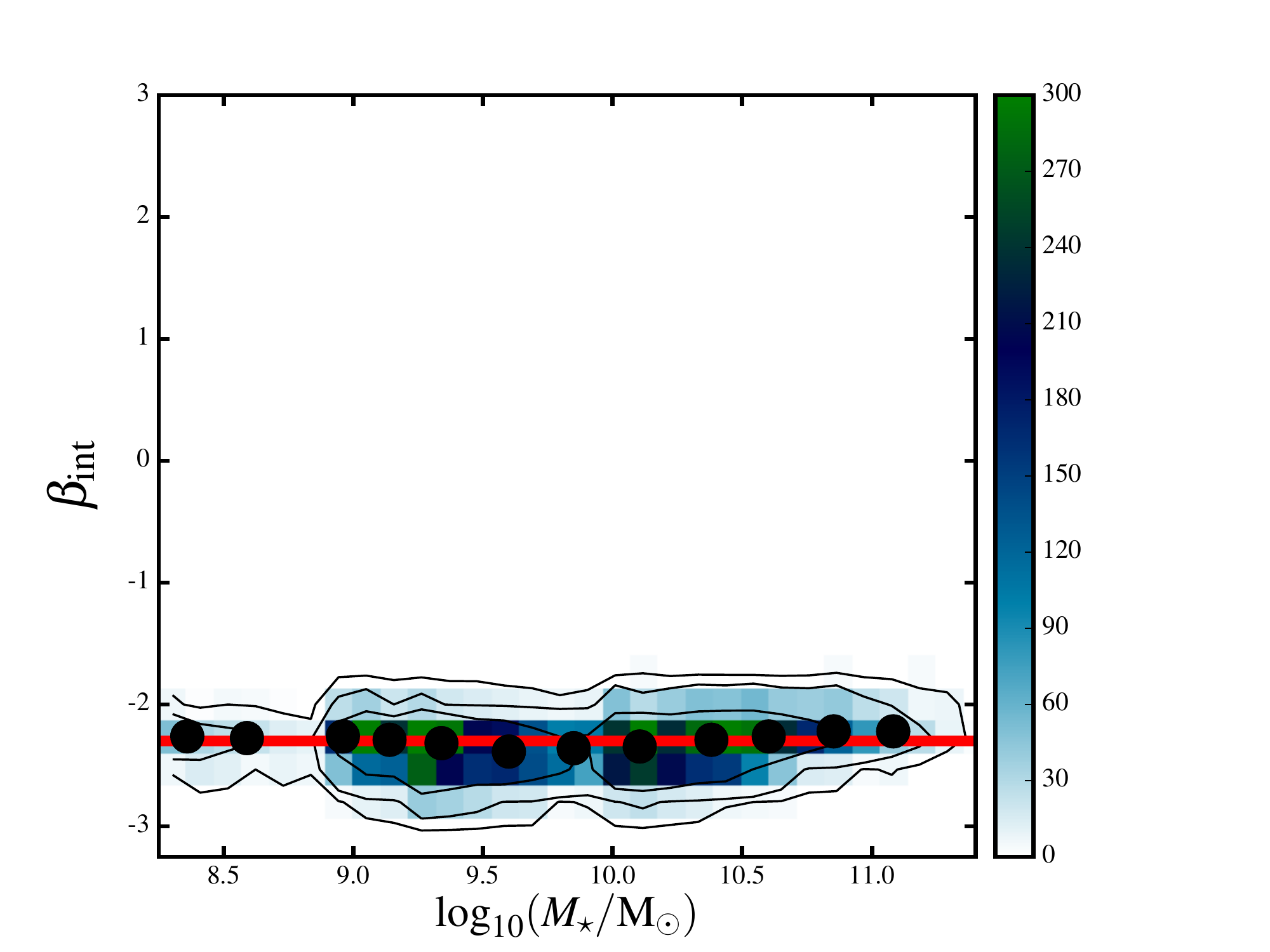}
\caption{The left-hand panel shows observed UV spectral slope
  ($\beta_{\rm obs}$) versus stellar mass for our
  mass-complete sample of $2.0<z<3.0$ star-forming galaxies selected
  from the HUDF, CANDELS GOODS-S, CANDELS UDS and UVISTA surveys (see
  text for details). The background image shows a 2D histogram of the
 $\beta_{\rm obs}$--$M_{\ast}$ distribution. The black data-points show the
 median values of $\beta_{\rm obs}$ in stellar-mass bins of width 0.25
 dex (uncertainties are smaller than the data points). The red line
 shows the best-fitting polynomial (equation 16) to the
individual, error-weighted, galaxies (a fit to the binned data points
produces a virtually identical result). The right-hand panel shows
intrinsic UV slope ($\beta_{\rm int}$) versus stellar mass for the
same sample. The background image shows a 2D histogram of the
$\beta_{\rm int}$--$M_{\ast}$ distribution, while the black data points
again show the median values binned by stellar mass. The values of $\beta_{\rm int}$ have been measured from
the best-fitting SED templates, prior to the addition of dust attenuation. 
The red line shows a constant value of $\beta_{\rm int}=-2.3$,
which is consistent with the data over the full stellar-mass range.}
\end{figure*}

\begin{figure}
\includegraphics[trim=0.7cm 0.2cm 1.0cm 1.0cm, clip, width=0.5\textwidth]{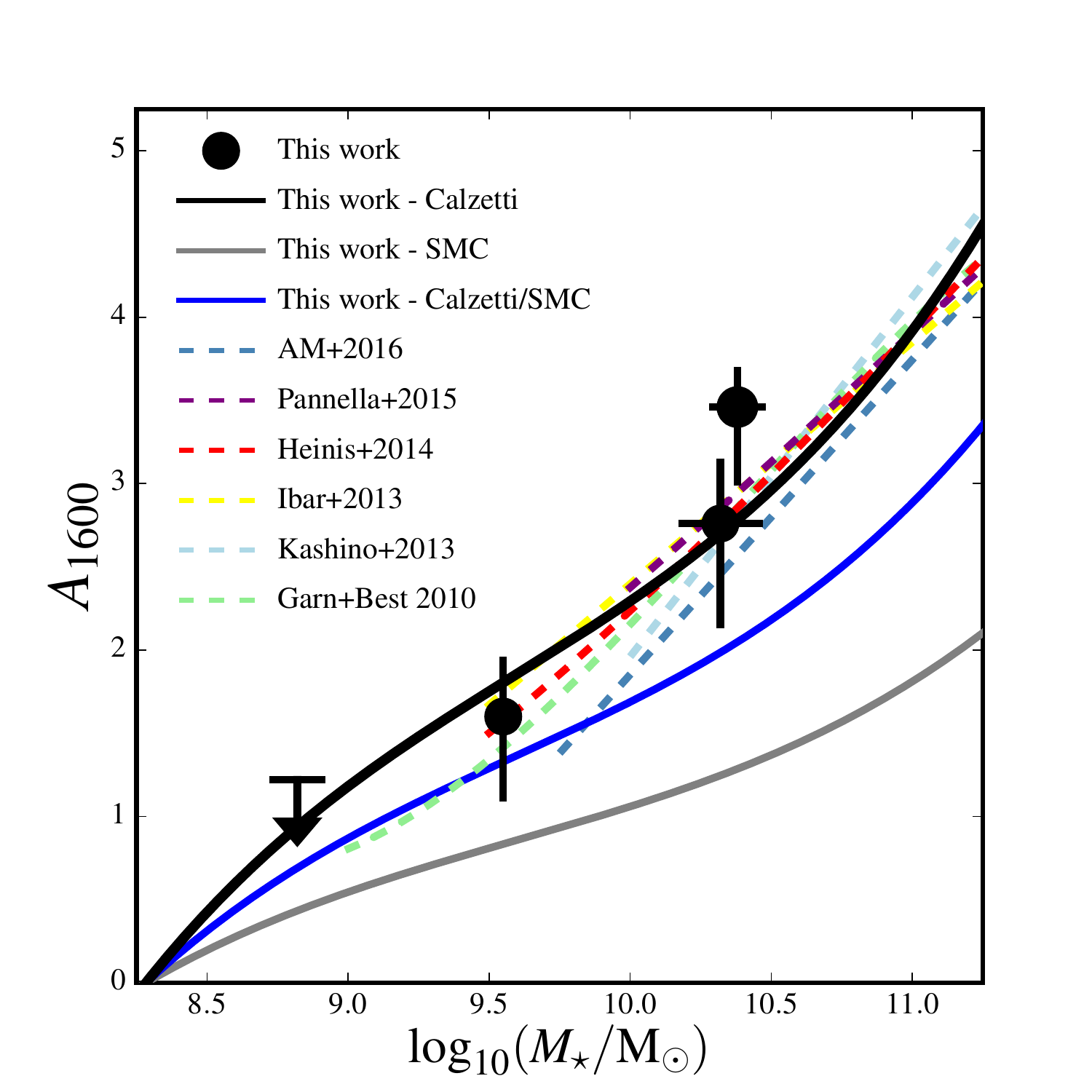}
\caption{UV attenuation ($A_{1600}$) versus stellar mass. The black
  data points show our estimates of $A_{1600}$ within the same
  stellar-mass bins shown in Figs.~1 and 3. The value of $A_{1600}$ in each
  mass bin has been calculated by converting between IRX and
  $A_{1600}$ using equation (7). The solid black and grey lines show predicted
$A_{1600}$--$M_{\ast}$ relations, based on combining our fit to the
$\beta$--$M_{\ast}$ relation (see Fig.~4a) with a Calzetti-like (black)
or SMC-like (grey) dust law. The solid blue line shows the predicted
$A_{1600}$--$M_{\ast}$ relation for a dust law with an intermediate UV
slope to the Calzetti and SMC-like dust laws (blue dashed line in Fig.~3).
The remaining dashed lines show $A_{1600}$--$M_{\ast}$ relations derived
from a number of different studies from the literature (see text for details).
These $A_{1600}$--$M_{\ast}$ relations have only been plotted over the mass range where they are properly
constrained.}
\end{figure}

\section{The relationship between stellar mass and UV attenuation}
Given that the UV attenuation ($A_{1600}$) is not directly observable,
in order to  study the $A_{1600}-M_{\ast}$ relation it is necessary
to employ an observational proxy. The standard
method for achieving this is to use IRX as the observational proxy for $A_{1600}$
(i.e. via equation 7). However, under the assumption that all star-forming
galaxies have the same, or at least very similar, {\it intrinsic} UV
slopes, it is also possible to use the observed UV slope ($\beta_{\rm  obs}$) as an observational proxy for $A_{1600}$.

In the recent literature there have been numerous studies that have
investigated the relationship between $\beta$ and absolute UV magnitude, the so-called
colour-magnitude relation (CMR), for Lyman break galaxies at $4 \leq z \leq 8$ (e.g. \citealt{sandy14}; \citealt{d13}; \citealt{fink12}; \citealt{bow14}).
The motivation behind these studies is to use the evolution of the CMR
to gain insight into the the relationship between UV luminosity, stellar population age/metallicity and dust reddening.
Although useful for constraining dust attenuation at high redshift (e.g. \citealt{ferg17}), the concentration on UV luminosity is largely
through necessity, with accurate stellar masses difficult to measure
at high redshift. In this section we investigate the relationship between stellar mass
and UV spectral slope for star-forming galaxies at  $2 \leq z \leq 3$,
taking advantage of the availability of deep {\it Spitzer} IRAC data, 
sampling rest-frame wavelengths of $\lambda_{\rm rest}\simeq 1\mu$m, to
deliver the necessary robust stellar-mass estimates.

\subsection{UV slope versus stellar mass}
The left-hand panel of Fig.~4 shows $\beta_{\rm obs}$ versus stellar mass for the
full sample of $UVJ-$selected star-forming galaxies at $2 \leq z \leq 3$. 
The underlying 2D histogram shows the distribution of the galaxies on
the $\beta_{\rm obs}$--$M_{\ast}$ plane, while the black data points show the
median values of $\beta_{\rm obs}$ in twelve $\Delta M_{\ast}=0.25$ dex mass
bins, spanning the full dynamic range of the sample.
The red curve shows the best-fitting third-order polynomial to the
error-weighted data, and has the following form:
\begin{equation}
\beta_{\rm obs} = -1.136 +0.589X + 0.130X^{2} + 0.106X^{3},
\end{equation}
\noindent
where $X=\log(M_{\ast}/10^{10}\Msun)$. A fit to the binned data
points produces a virtually identical result. The right-hand panel of Fig.~4 shows intrinsic UV-slope
($\beta_{\rm int}$) versus stellar mass, where $\beta_{\rm int}$ has been 
measured from the best-fitting SED templates, before the application of dust reddening.
It can be seen that the distribution of intrinsic UV slopes is fully
consistent with a constant value of $\beta_{\rm int}=-2.3\pm 0.15$
over the full range in stellar mass, where the quoted uncertainty is
the standard deviation rather than the standard error. 

It should be noted that the relatively small scatter around $\beta_{\rm int}=-2.3$ is not an inevitable consequence of our choice
of SED templates. Indeed, the templates used to measure the intrinsic UV slopes, 
which included reddening based on both the \cite{calz00} and SMC dust curves, cover the range $-2.75 \leq
\beta_{\rm int} \leq -1.55$. Moreover, we recovered a virtually
identical result for the median value of $\beta_{\rm int}$ when we
re-fitted the star-forming galaxies in the HUDF with low-metallicity
BPASSv2 stellar population models, including the contribution from
binary stars and nebular line and continuum emission (see Section 7.2).

Fundamentally, the $\beta_{\rm int}$ results shown in the right-hand
panel of Fig.~4 provide reassurance that the increase in
$\beta_{\rm obs}$ shown in the left-hand panel can be attributed to
increasing dust attenuation, rather than increasing stellar population
age/metallicity.

\subsection{The $A_{1600}$--$M_{\ast}$ relation}
The immediate utility of equation (16) is that, for a given assumption
about the form of the attenuation law in the UV (equation 12) and the
value of $\beta_{\rm int}$, it provides a
direct prediction of the form of the $A_{1600}$--$M_{\ast}$ relation. For example, for the \cite{calz00} attenuation law and $\beta_{\rm  int}=-2.3$:
\begin{equation}
A_{1600}=2.293 +1.160X + 0.256X^{2} + 0.209X^{3},
\end{equation}
\noindent
where $X=\log(M_{\ast}/10^{10}\Msun)$. This relation is shown as the
solid black line in Fig.~5, along with the equivalent relations for
the \cite{gord03} SMC extinction law (solid grey line) and a dust law
with an intermediate UV slope to the Calzetti and SMC dust laws
(solid blue line), all with $\beta_{\rm int}=-2.3$. In addition, the black data points show the results of our stacking
analysis, where we have converted between IRX and $A_{1600}$ using
equation (7). In Fig.~5 we also plot $A_{1600}$--$M_{\ast}$ relations
derived from the results of six different literature studies. Five of
these studies are focused on the properties of star-forming galaxies in the redshift range
$1.5<z<3.0$. The final study defines the local $A_{1600}$--$M_{\ast}$ relation,
and is based on a large sample of star-forming galaxies drawn from the SDSS \citep{garn10}.

\cite{am16} provided binned measurements of $A_{\rm FUV}$ as a function of
stellar mass for a large statistical sample of $2.5<z<3.5$ LBGs in
the COSMOS field. The definition of $A_{\rm FUV}$ used by \cite{am16} is 
virtually identical to our definition of $A_{1600}$ (i.e. equation 7) and 
consequently, the blue dashed curve in Fig.~5 shows our
linear fit to their binned $A_{\rm FUV}$ data.
\cite{heinis14} provide a relation between IRX and stellar mass at $z\simeq 1.5$, based on their study of UV-selected galaxies in COSMOS.
To produce the red curve shown in Fig.~5 we have converted
their IRX$-M_{\ast}$ relation into a $A_{1600}$--$M_{\ast}$ relation
using equation (7).

\cite{pan15} investigated the dust attenuation properties of a large 
sample of $UVJ-$selected star-forming galaxies at $z\leq 4$ in the
GOODS-N field, based on stacking the ultra-deep {\it Herschel} PACS and
SPIRE data. To plot the purple dashed curve shown in Fig.~5, we have used their $A_{\rm UV}$--$M_{\ast}$ relation at $z\simeq 2.3$ and
corrected for the difference between their definition of $A_{\rm UV}$ and our definition of $A_{1600}$.
We have also decreased the \cite{pan15} stellar-mass estimates by
$0.22$ dex in order to convert from a \cite{salp55}  to a \cite{chab03} IMF. 

\begin{figure*}
\includegraphics[width=15cm]{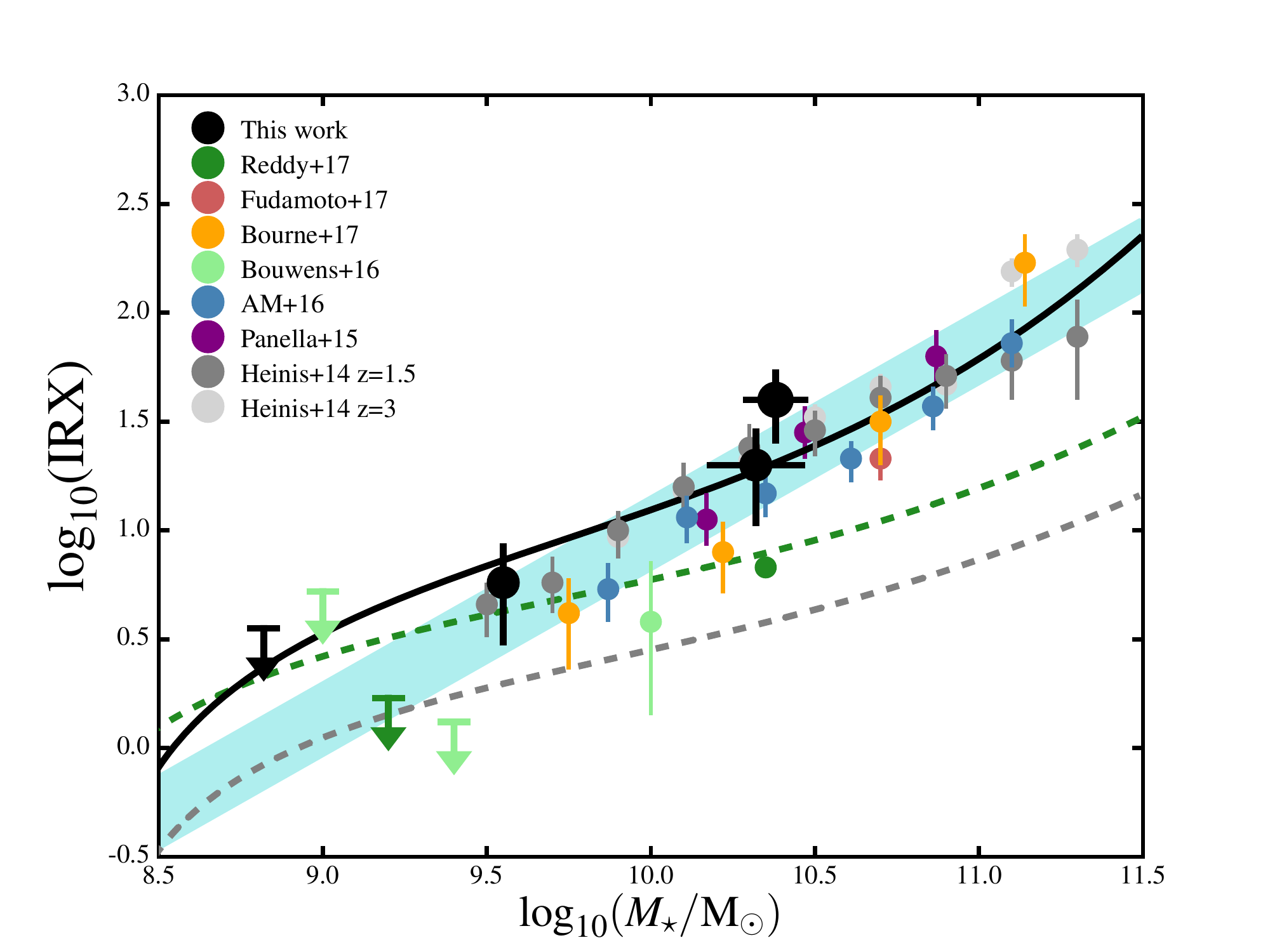}
\caption{The IRX--$M_{\ast}$ relation for star-forming galaxies at
  $z\simeq 2.5$. In addition to our new results, based on stacking the
  1.3-mm ALMA data within the HUDF (black data points), we also plot results derived
  by \protect\cite{Reddy17}, \protect\cite{fuda17},
  \protect\cite{nathan17}, \protect\cite{bow16}, \protect\cite{am16},
  \protect\cite{pan15} and \protect\cite{heinis14}.
The solid black and dashed grey lines show predicted IRX$-M_{\ast}$
relations based on combining our fit to the $\beta$--$M_{\ast}$
relation (see Fig.~4a) with the IRX$-\beta$ relations based on either
a Calzetti-like (black) or SMC-like (grey) dust law (assuming
$\beta_{\rm int}=-2.3$). The dark green dashed line is the equivalent  IRX$-M_{\ast}$ prediction
based on the SMC-like IRX-$\beta$ relation derived by
\protect\cite{Reddy17}, which assumes $\beta_{\rm int}=-2.62$.
The shaded blue region represents our linear $\log({\rm
  IRX})-\log(M_{\ast}/\Msun)$ fit to the combined data set and its associated
$1\sigma$ scatter.}
\end{figure*}

\subsubsection{Conversion from nebular to continuum attenuation}
The other three $A_{1600}$--$M_{\ast}$ relations plotted in
Fig.~5 are derived from studies which originally provided
a measurement of the nebular attenuation experienced by the H$\alpha$
emission line ($A_{\rm{H}\alpha}$). \cite{garn10} and \cite{kash13} measured
  $A_{\rm{H}\alpha}$ via the Balmer decrement for galaxies at
  $z\simeq0$ and $z\simeq 1.5$, respectively, while \cite{ibar13} obtained a measurement of
$A_{\rm{H}\alpha}$ at $z\simeq 1.5$ by matching observed H$\alpha$ 
and integrated IR luminosities.

To plot these relations in Fig.~5 it is necessary to both convert between
attenuation at $\lambda_{\rm rest}=6563\AA$ and $\lambda_{\rm rest}=1600\AA$, and to estimate the ratio of nebular-to-continuum attenuation in $z\simeq 2.5$ star-forming galaxies.
When deriving their stellar attenuation curve from local starburst
galaxies, \cite{calz94} determined that
$E(B-V)_{\rm star}=0.44E(B-V)_{\rm neb}$, based on the assumption that
the nebular component follows a Galactic extinction curve
(i.e. $R_{V}~=~3.1$) and that the stellar continuum component followed the \cite{calz94} attenuation law. 
Under this prescription the \cite{calz94} result indicates that $A_{\rm star}\simeq 0.59\times A_{\rm neb}$ at $\lambda_{\rm rest}~=~6563\AA$ \citep{calz01}.

Whether or not a significant difference between stellar continuum and
nebular reddening persists in high-redshift star-forming
galaxies is still a matter of debate (e.g. \citealt{wuyts13}; \citealt{kash13}; \citealt{price14}).
Here we adopt the results of \cite{kash13}, who investigated nebular versus stellar attenuation in a sample of 89 star-forming galaxies at
$1.4<z<1.7$, using FMOS near-IR spectra. Assuming that both nebular and
continuum attenuation followed the \cite{calz00} law, \cite{kash13} 
found that $E(B-V)_{\rm star}=f\times E(B-V)_{\rm neb}$, with $f$ in
the range $0.69< f < 0.83$. Based on these results we adopt a value of
$f=0.76$, noting that this is somewhat higher than the value of
$f=0.54_{-0.10}^{+0.13}$ derived for $z\simeq 1.4$ star-forming galaxies by \cite{price14}.

Consequently, in Fig.~5 the $A_{1600}$--$M_{\ast}$
relations for \cite{garn10}, \cite{ibar13} and \cite{kash13} 
were derived by first multiplying their determinations of nebular attenuation
by $f=0.76$ to obtain an estimate of the continuum attenuation at
$\lambda_{\rm rest}=6563\AA$,  before assuming that $A_{1600}=3\times A_{6563}$, as appropriate for the \cite{calz00} attenuation law.
As part of this calculation, the stellar-mass estimates provided by \cite{ibar13} and
\cite{kash13} were reduced by $0.22$ dex to convert from 
a \cite{salp55} to a \cite{chab03} IMF and the stellar-mass estimates
of \cite{garn10} were reduced by $0.03$ dex to convert from a \cite{kroupa01} IMF to a \cite{chab03} IMF. 
The $A_{1600}$--$M_{\ast}$ relations plotted in Fig.~5 for \cite{garn10}
and \cite{ibar13} are based on our corrected versions of the
functional forms they presented for the
$A_{\rm{H}\alpha}$--$M_{\ast}$ relation. In contrast, the dashed curve
shown in Fig.~5 for \cite{kash13} is our linear fit to corrected
versions of their binned $A_{\rm{H}\alpha}$ values as a function of stellar mass.

After making the corrections between nebular attenuation ($A_{{\rm H}\alpha}$) and stellar continuum attenuation
in the UV ($A_{1600}$), the results of \cite{garn10}, \cite{ibar13} and \cite{kash13} are in good
agreement with the more direct determinations of \cite{am16},
\cite{pan15} and \cite{heinis14} \footnote{This would not
  be true based on the SMC extinction curve,
which predicts $A_{1600}$ values approximately two times higher ($A_{1600}=5.5\times A_{6563}$).}.
Notably, despite the wide variety of different data sets, redshift ranges
and methodologies represented, there is actually remarkably good
agreement on the form of the $A_{1600}$--$M_{\ast}$ relation,
particularly at high stellar mass.  
The results of the different studies shown in Fig.~5 indicate that
galaxies with $M_{\ast}\simeq10^{10}\Msun$ 
experience $A_{1600}\simeq 2$ magnitudes of UV attenuation on average, 
which rises to $A_{1600}\simeq 4$ magnitudes of attenuation at $M_{\ast}\simeq 10^{11}\Msun$. 
It can clearly be seen from Fig.~5 that both our stacking
results (black data points) and the predicted $A_{1600}$~--~$M_{\ast}$
relation based on combining the $\beta$--$M_{\ast}$ relation presented
in Fig.~4a with the \cite{calz00} attenuation law (solid black curve) are fully consistent with this consensus. 

Finally, it is also clear from Fig.~5 that the $A_{1600}$--$M_{\ast}$
relations predicted by both an SMC-like dust law (grey solid line) and 
a dust law with a UV slope intermediate to the SMC and Calzetti dust
laws (solid blue line) are inconsistent with our new results and those
in the literature. Note that this conclusion regarding the SMC-like dust law is not significantly affected by our assumption that
$\beta_{\rm int}~=~-2.3$, since even shifting to $\beta_{\rm int}=-2.6$ only increases
the vertical normalization of the SMC-like relation by $\Delta A_{1600}\simeq 0.3$ mag. 

\section{The IRX$-M_{\ast}$ relation}
In Fig.~6 we show the IRX$-M_{\ast}$ relation for
star-forming galaxies at $z \simeq 2.5$. In addition to our new
results we also plot the results derived by a number of other 
studies, the details of which have been previously described in
Sections 4 and 5.

Given the growing consensus in the literature that it is 
fundamentally stellar mass that drives UV attenuation, and therefore 
detectability at FIR, sub-mm and mm-wavelengths
(e.g. \citealt{heinis13}; \citealt{pan15}; \citealt{am16}; \citealt{d17}; \citealt{bow16}; \citealt{Reddy17}), it is
arguable that the IRX$-M_{\ast}$ relation is physically more interesting than the IRX$-\beta$ relation. 
Excluding the most heavily obscured sources, the direct mapping
between IRX and $A_{1600}$ means that both relations provide a useful
method for estimating a physical quantity ($A_{1600}$), which is not directly observable.

Within this context, the IRX$-\beta$ relation has the advantage of
being based on a straightforward observable ($\beta$), 
in contrast to the IRX$-M_{\ast}$ relation, which is based on a
derived quantity, namely the stellar mass.
However, the much shallower form of the IRX$-M_{\ast}$ relation means that, in reality, it is the derived stellar mass that is likely
to provide the cleanest prediction of UV attenuation. 

An illustration of this is provided by considering a star-forming galaxy with
$\log(M_{\ast}/\Msun)=10.0$ and $\beta=-1.15$ (see
Fig.~4a). According to both the IRX$-\beta$ and
IRX$-M_{\ast}$ relations based on Calzetti-like attenuation (black curves in Fig.~1 and
Fig.~6) this galaxy should have $\log(\rm IRX)~=~1.1$. However,
if we assume that representative uncertainties on the
$\beta$ and $\log(M_{\ast}/\Msun)$ measurements are $\pm 0.3$ and
$\pm 0.3$ dex respectively, the IRX$-\beta$ relation implies that
$\log(\rm IRX)=1.1\pm0.3$, whereas the IRX$-M_{\ast}$ relation
implies that $\log(\rm IRX)=1.1\pm0.15$. 
As a consequence, the estimate of UV attenuation derived from the IRX$-M_{\ast}$ relation carries half the
uncertainty of that derived from the IRX$-\beta$ relation
(i.e. $A_{1600}=2.3\pm0.3$ mag compared to $A_{1600}=2.3\pm
0.6$ mag).

It is clear from Fig.~6 that our predicted IRX-$M_{\ast}$
relation based on Calzetti-like dust attenuation (solid black curve)
is in excellent agreement with the combined literature data set at
$\log(M_{\ast}/\Msun)\geq 10$.
In the mass interval $9.5<\log(M_{\ast}/\Msun)<10$ the predicted
IRX$-M_{\ast}$ relation is in good agreement with our stacked data
point at $\log(M_{\ast}/\Msun)=9.6$ and the data from
\cite{heinis14}, but in general appears to overpredict the combined literature data by $\simeq 0.2$ dex.

In contrast, it is clear that the IRX$-M_{\ast}$ relation predicted by
an SMC-like extinction law with $\beta_{\rm int}=-2.3$ (grey dashed
line) is a very poor match to the literature data, underpredicting
the observed IRX values by $\simeq 0.4$ dex at
$\log(M_{\ast}/\Msun)\simeq 9.5$ and by $\simeq 1.0$ dex at $\log(M_{\ast}/\Msun)\simeq 11.0$.
The situation is improved somewhat if, following \cite{Reddy17}, the SMC-like prediction
is changed to assume that $\beta_{\rm int}=-2.62$ (green dashed line). In this case, the SMC-like IRX$-M_{\ast}$ relation does a
good job of reproducing the literature data at
$\log(M_{\ast}/\Msun)\simeq 9.5$. However, it can
be seen that the SMC-like relation still does a very poor job of reproducing the
literature data at $\log(M_{\ast}/\Msun)\geq 10$, underpredicting
the observed IRX values by $\simeq 0.6$ dex at $\log(M_{\ast}/\Msun)\simeq 11$.

The blue shaded area in Fig.~6 represents our best-fitting linear relationship between $\log(M_{\ast}/\Msun)$ and
$\log(\rm IRX)$ for those data-points at $\log(M_{\ast}/\Msun)\geq 9.5$. This has the following form:

\begin{equation}
\log(\rm IRX) = 0.85(\pm 0.05)\log\left(\frac{M_{\ast}}{10^{10}\Msun}\right) + 0.99(\pm 0.03),
\end{equation}

where the width of the shaded region illustrates the $1\sigma$ scatter
associated with the best-fitting relation ($\simeq 0.17$ dex). It can be seen from
Fig.~6 that this linear relation provides an excellent
description of the combined data set, such that a more
complicated functional form is not justified. We note that our
best-fitting relation is very similar to that previously derived by
\cite{am16} at $z\simeq 3$ and somewhat steeper than, but still consistent with, the relation derived
by \cite{heinis14} at $z\simeq1.5$. Finally, we note that our
best-fitting relation also provides a good description of the
IRX$-M_{\ast}$ data for $1.5<z<2.5$ star-forming galaxies derived from
the 3D-HST photometric catalogues by \cite{whit14}, using $L_{\rm
  IR}$ values based on {\it Spitzer} 24-$\mu$m data.

In general, it can be seen from Fig.~6 that the IRX$-M_{\ast}$ relations 
based on SMC-like extinction curves (grey and green dashed lines) have fundamentally the wrong shape
to provide a good description of the literature data over a wide
range in stellar mass. This is equivalent to saying that the slope
of the effective attenuation curve needs to be closer to
$dA_{\mbox{\tiny 1600}}/d\beta \simeq 2$ than $dA_{\mbox{\tiny
    1600}}/d\beta \simeq 1$ in order to reproduce the observed IRX$-M_{\ast}$ data. Therefore, it appears 
that the effective dust attenuation experienced by
star-forming galaxies at $z\simeq 2.5$ is much more similar to the
\cite{calz00} attenuation law than an SMC-like extinction curve, at least at $\log(M_{\ast}/\Msun)\geq 9.75$. 

Given the sensitivity limits of currently available data, it is clear
from Figs.~5 and 6 that the form of the dust attenuation law at
$\log(M_{\ast}/\Msun)\leq 9.75$ is still very poorly constrained. 
However, given the constraints that do exist, it appears likely that
the IRX$-M_{\ast}$ relation will fall below the Calzetti-like
prediction at $\log(M_{\ast}/\Msun)\simeq 9.0$. If confirmed, this 
would indicate that there is a shift towards a steeper attenuation law with decreasing stellar mass, something which could
plausibly be connected to a mass-dependent change in the gas/dust geometry
(e.g. \citealt{koop15}). ALMA imaging at sub-mm wavelengths with an effective dust continuum
sensitivity a factor 3--5 deeper than the 1.3-mm mosaic
exploited here could potentially address this question.

\subsection{Star-formation rate density}
Armed with our best-fitting IRX$-M_{\ast}$ relation (equation 18) it is possible to estimate the total (UV+IR) star-formation rate density at
$z\simeq 2.5$ using our sample of $2<z<3$ star-forming galaxies in the
HUDF. For each galaxy within the HUDF sample with stellar-mass $\log(M_{\ast}/\Msun)\geq
8.5$, we calculate the SFR$_{\rm TOT}$ (UV+IR) assuming that
\begin{equation}
{\rm SFR}_{\rm UV+IR} = {\rm SFR}_{\rm UV}(1+{\rm IRX}),
\end{equation}
where SFR$_{\rm UV}$ and IRX are calculated from equation (10) and
equation (18) respectively, and $L_{\rm UV}$ and $M_{\ast}$ are derived from the 
best-fitting SED template to its UV-to-midIR photometry. By simply summing the
contribution of each galaxy, and taking the co-moving volume of the HUDF to be $1.5\times 10^{4}$
Mpc$^{3}$, we arrive at an estimate of $\log(\rho_{\rm  UV+IR}/\Msun$yr$^{-1}$Mpc$^{-3})=-1.04\pm{0.09}$, where the
uncertainty has been calculated using bootstrap re-sampling. This
estimate is in excellent agreement with the value calculated by \cite{d17} and the $\rho_{\rm
  SFR}(z)$ fitting formula derived by
\cite{beh13}. Our estimate is somewhat higher than the $\rho_{\rm SFR}(z)$ fitting formula
derived by \cite{mad14}, but is actually very consistent with their compilation of data at $z\simeq 2.5$.
If we also account for the fact that, in the HUDF at least, the best-fitting IRX$-M_{\ast}$
relation underpredicts the SFR$_{\rm IR}$ values of the most heavily obscured sources by a factor of $\simeq 2$, 
our SFRD estimate increases to $\log(\rho_{\rm  UV+IR}/\Msun$yr$^{-1}$Mpc$^{-3})=-0.94\pm{0.13}$.

\section{Discussion}
\begin{figure*}
\includegraphics[trim=0.5cm 0.05cm 3.0cm 1.0cm, clip, width=0.48\textwidth]{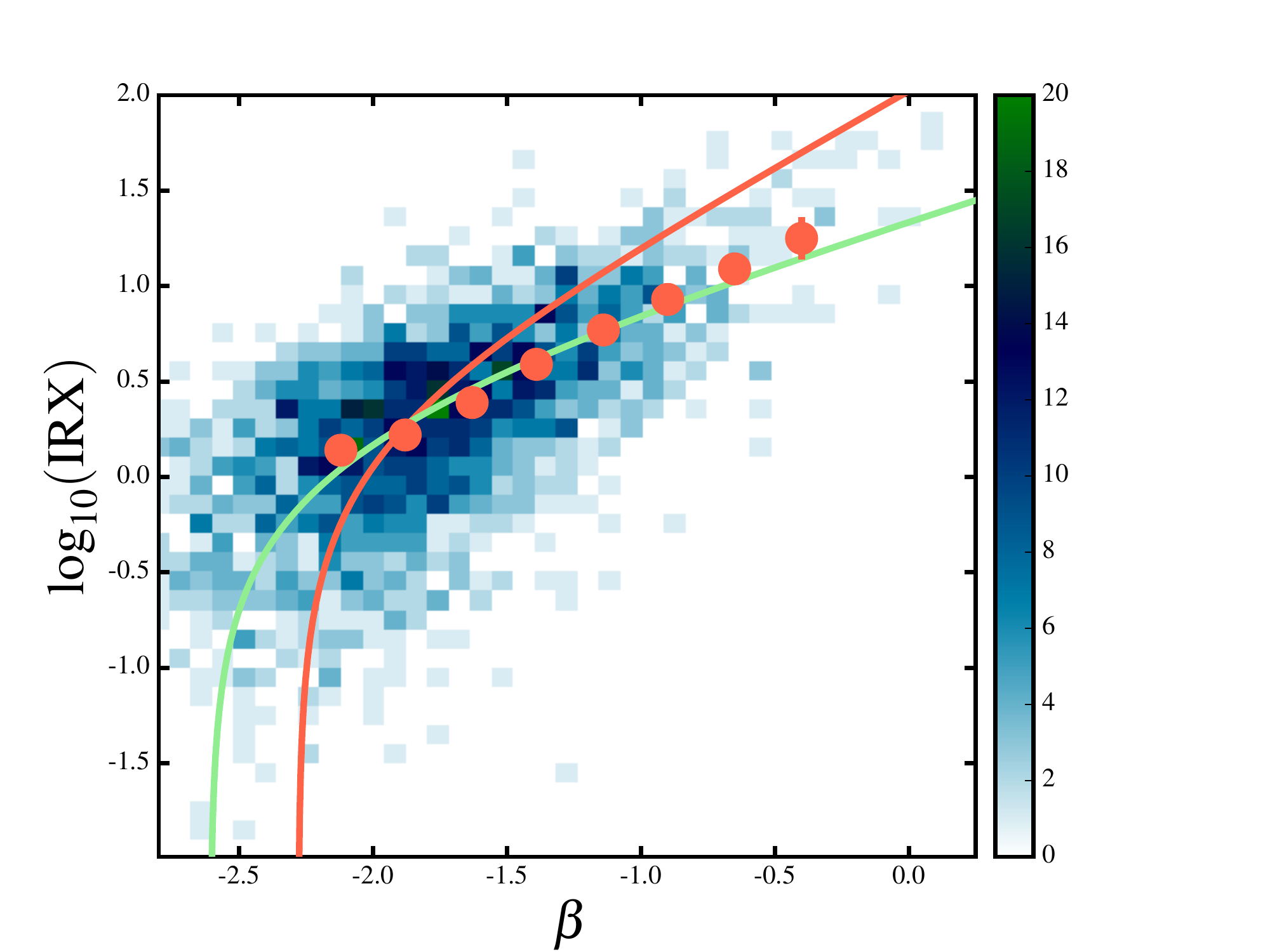}
\includegraphics[trim=0.5cm 0.05cm 3.0cm 1.0cm, clip, width=0.48\textwidth]{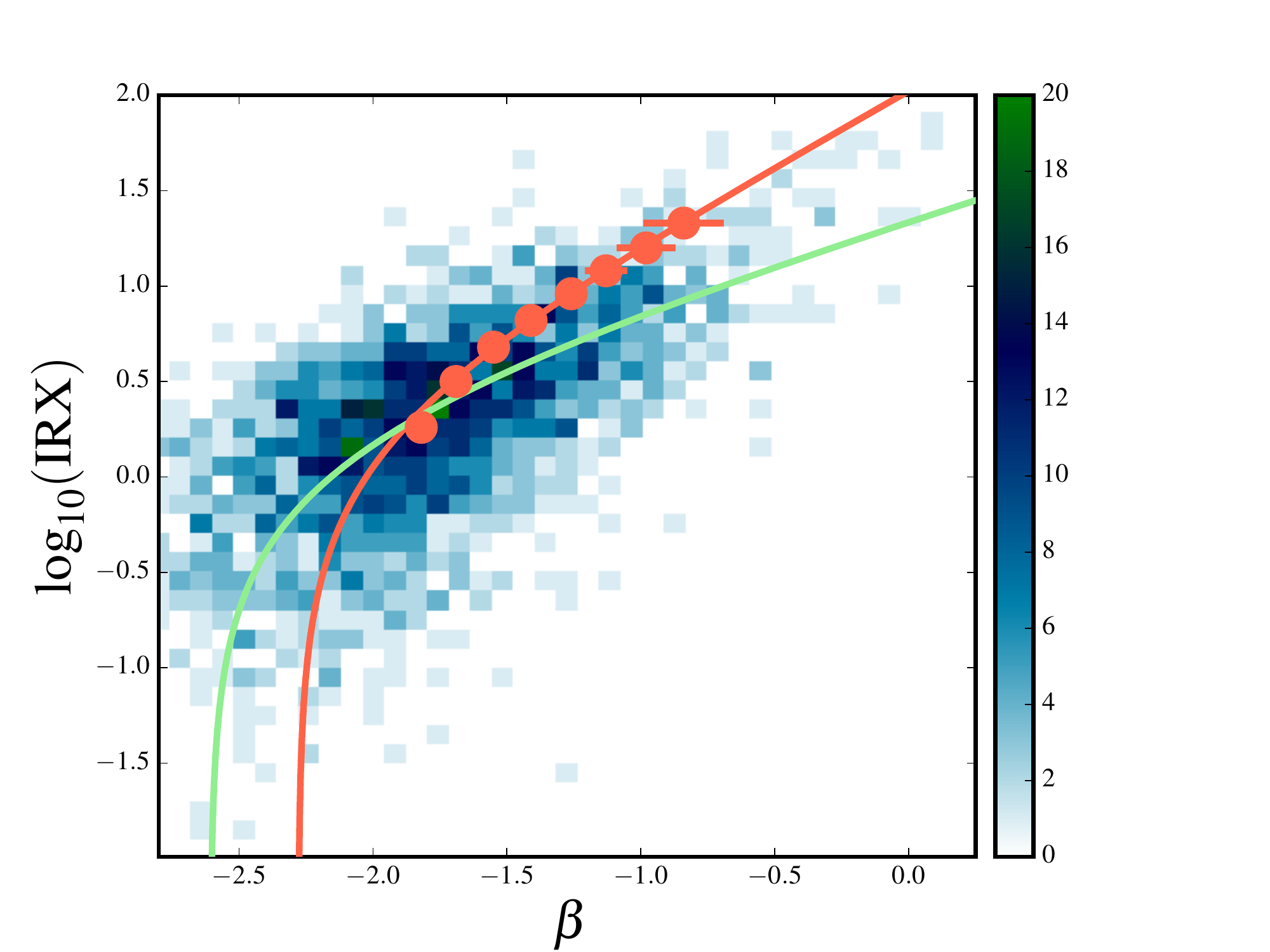}
\caption{Simulations of the IRX$-\beta$ relation. Both panels show a 2D histogram of a single realisation of
  our IRX$-\beta$ simulation, along with the input IRX$-\beta$ relation (red line) and 
the IRX$-\beta$ relation recently determined for $1.5<z<2.5$ star-forming galaxies (green line) by \protect\cite{Reddy17}. 
Each realisation of the simulation features 3545 simulated galaxies
to match the size of the observed sample of \protect\cite{Reddy17}. In the
left-hand panel, the red data points show the mean binned values of IRX determined from 1000 realisations, when binning the
simulation output in terms of $\beta$. The eight data points
correspond to $\Delta\beta=0.25$ 
bins spanning the range $-2.25 \leq \beta \leq -0.25$.
In the right-hand panel the red data points show the equivalent
results when the output is binned in terms of stellar mass. 
The eight data points shown in the right-hand panel correspond (left-to-right) to the
stellar-mass range $8.75\leq \log(M_{\ast}/\Msun)\leq 10.75$, in bins of
width 0.25 dex. It can be seen that the IRX$-\beta$ relationship recovered when
binning in terms of $\beta$ is biased, a problem that does not occur when
binning in terms of stellar mass.}
\end{figure*}

The results presented in the last two sections strongly suggest that
the effective attenuation law experienced by star-forming galaxies
at $z\simeq 2.5$ lies closer to the \cite{calz00} attenuation law than
the SMC-like extinction law.

As discussed in Section 1, it is expected that the attenuation law applicable to integrated galaxy
photometry should be substantially greyer than a SMC-like extinction law if the local extinction law {\it is} SMC-like.
Although it is impossible to directly measure the extinction law in $2<z<3$ star-forming galaxies, supporting evidence for
this scenario is provided by the direct extinction curve measurements from gamma-ray burst (GRB) afterglows. These studies
typically show that GRB extinction curves are similar to the SMC
extinction curve (e.g. \citealt{grb}), at least at low-to-moderate levels of extinction (i.e. A$_{\rm V}\leq 0.65$), exactly as
required to produce a Calzetti-like attenuation curve.

However, despite the consistency with the results presented here, it is clear from Fig.~3 
that the majority of studies in the literature find that the
IRX$-\beta$ relation displayed by star-forming galaxies at this epoch
falls significantly below what would be expected for the \cite{calz00}
attenuation law. This presents a significant problem, given that it is
clearly necessary to have consistency between the IRX$-\beta$ and IRX$-M_{\ast}$ relations.

In Section 4 we proposed three potential solutions to the apparent
disagreement between the IRX$-\beta$ relations determined by various
literature studies and the IRX$-\beta$ relation expected if the underlying attenuation law was Calzetti-like.
The first, and most straightforward, solution is to accept that the 
effective attenuation law has an intermediate dust slope of $dA_{\mbox{\tiny 1600}}/d\beta \simeq 1.45$. However, although this
is able to do a reasonable job of reproducing the observed IRX$-\beta$
data, it is unable to reproduce the observed $A_{1600}$~--~$M_{\ast}$ or IRX$-M_{\ast}$ relations (see Fig.~5).

The second option is to adopt an SMC-like dust law in combination
with an ultra-blue value of intrinsic UV slope (i.e. $\beta_{\rm
  int}=-2.62$), as proposed by \cite{Reddy17}. However, as shown in Fig.~6, the IRX$-M_{\ast}$
relation predicted by an SMC-like dust law, regardless of the
assumed value of $\beta_{\rm int}$, does not have the correct shape to
reproduce the observed data over a large dynamic range in stellar mass.

The third and final option proposed in Section 4 is that the
underlying dust law at $z\simeq 2.5$ is indeed close to
the \cite{calz00} attenuation law, and that the IRX$-\beta$ relations derived by
previous studies are biased towards low values of IRX at red values of $\beta$.
In this scenario, the bias in the IRX$-\beta$ results is directly 
caused by uncertainties in measuring individual values of $\beta$, in
combination with using mean IRX values binned by $\beta$.
In this section we investigate this remaining option in more detail.

\subsection{Bias due to UV slope uncertainty}
The IRX-$\beta$ plot is vulnerable to bias
when plotting results binned by $\beta$ due to a
combination of three elements: the relatively large uncertainties associated with observational determinations of $\beta$, the steepness
of the stellar-mass function and the steepness of the IRX$-\beta$ relationship itself. 

The steepness of the stellar-mass function means that any given sample of
star-forming galaxies must be dominated by galaxies close to the
stellar-mass limit of the sample. Moreover, due to the strong
correlation between stellar mass and dust attenuation, these low-mass 
galaxies will also have the bluest values of $\beta$ and the smallest values of IRX. 
Furthermore, due to the large uncertainties inherent in measuring
$\beta$ photometrically, even relatively modest amounts of
photometric scatter will result in a small fraction of these objects
being measured as having much redder values of $\beta$. 
Unfortunately, due to the dominance of intrinsically blue objects ensured by the stellar-mass function,
even a small fraction of misclassified blue galaxies can lead to
significant problems when choosing to bin-up the IRX$-\beta$ plane in terms of the {\it observed} values of $\beta$. 

As a specific example, when considering the typical value of IRX for galaxies with observed values of $\beta$ in the range
$-1.5~<~\beta~<~-1.0$, this $\beta$ bin will inevitably contain significant
numbers of galaxies whose actual value of observed $\beta$ (i.e. with
perfect photometry) should be $\beta\simeq -2$. Given the steepness of
the IRX$-\beta$ relationship, these contaminants will carry 
significantly smaller values of IRX (factor of $\simeq 4$ for
Calzetti-like dust attenuation) than suggested by their measured UV-slope of $\beta\simeq
-1.25$. Fundamentally, it is this contamination from objects scattered
towards redder values of $\beta$ which inevitably results in the
apparent suppression of the average IRX values when binning samples by
observed $\beta$. Crucially, as explained in the previous section, the shallower slope
of the IRX$-M_{\ast}$ relation means that this bias should not apply
to determinations of the IRX$-\beta$ relation using samples binned by stellar mass.

\subsection{IRX simulation}
To investigate this problem further, we have run a simple simulation to
investigate whether IRX$-\beta$ results in the recent literature can
be reproduced by modelling the bias discussed above. We have focused
this simulation on reproducing the results of \cite{Reddy17}, but the 
same arguments apply to any of the studies discussed in Section 4.
The aim of this experiment is simply to ascertain whether or not an IRX$-\beta$ relation that is apparently
consistent with an SMC-like dust law and ultra-blue intrinsic UV
slope, can in fact be generated from a galaxy sample that is reddened
by the \cite{calz00} attenuation law.

The first step in this process is to define a fiducial intrinsic SED template for each of the synthetic galaxies.
Following \cite{Reddy17}, we fit the SEDs of the objects within our HUDF galaxy sample with constant star-formation
rate (CSFR) stellar population models based on a {\tt BPASSv2-100bin-0.002}
simple stellar population (\citealt{kroupa01} IMF with a 100$\Msun$
cut-off, $Z=0.002$\footnote{This is 0.1$Z_{\odot}$ for
  $Z_{\odot}=0.02$ or $0.14Z_{\odot}$ for $Z_{\odot}=0.014$
  \citep{asp09}.} and binary stars included), with accompanying nebular continuum and line emission predictions
from {\sc cloudy} \citep{fer13}. Full details about the production of
these models can be found in \cite{ferg17}. 
Based on the fits to our HUDF sample, the typical low-mass star-forming galaxy at $z\simeq 2.5$
is best-fit by a CSFR model with an age of 400 Myr (mass-weighted age
of 200 Myr) and an intrinsic UV slope of $\beta_{\rm int}=-2.28$. This
is therefore the fiducial intrinsic SED assumed for every synthetic galaxy. We note
that the typical best-fitting template based on the same models without the nebular continuum has an intrinsic UV slope of $\beta_{\rm int}=-2.48$.

After deciding on a fiducial SED, the next step is to draw a synthetic population of $z=2.5$
galaxies from the stellar-mass function of star-forming galaxies at
this redshift, which is assumed to follow a Schechter function with
the following parameters: $\log(\phi^{\ast}/{\rm Mpc}^{-3})=-3.9$, $\log(M_{\ast}/\Msun)=11.0$,
$\alpha=-1.7$. These parameters are consistent with the results of
both \cite{tom14} and \cite{david17}. Each synthetic galaxy was then allocated a value of dust attenuation ($A_{1600}$), based on our prediction for the
$A_{1600}$--$M_{\ast}$ relation (black curve in Fig.~5), with
an assumed Gaussian scatter of $\sigma=0.3$ mag (\citealt{garn10}).
The value of IRX for each synthetic galaxy could then be calculated
via equation (7).

According to the properties of the fiducial SED, and our modelling of
the full $2<z<3$ star-forming galaxy sample, the value of intrinsic
UV-slope ($\beta_{\rm int}$) associated with each 
synthetic galaxy was randomly drawn from a Gaussian distribution
centred on $\beta_{\rm int}=-2.28$ with $\sigma=0.15$. 
The exact value of the observed UV-slope was then calculated by applying the \cite{calz00} attenuation law with
the appropriate $A_{1600}$ value. The actual observed value of the
UV-slope (i.e. the measured value) was then calculated by applying a measurement uncertainty ($\sigma_{\beta}$) designed to reflect the
impact of photometric scatter. The value of $\sigma_{\beta}$ is a free
parameter which was varied in order to determine the level of $\beta$
uncertainty required to reproduce the \cite{Reddy17} results.

\subsubsection{Simulation output}
The output of a single realisation of our simulation is shown in
Fig.~7, where we have applied the $H<27$ and
$m_{1700\times(1+z)}<27$ cuts adopted by \cite{Reddy17} and applied a
constant value of $\sigma_{\beta}=0.35$. Both panels show the same  2D histogram of the simulation output on
the IRX$-\beta$ plane, with the red and green curves showing the input
IRX$-\beta$ relation and the observed relation from \cite{Reddy17} respectively.
In the left-hand panel, the red data points show the mean binned
values of IRX produced by 1000 realisations, binning in $\beta$.
The eight data points correspond to $\Delta\beta=0.25$ bins spanning
the range $-2.25 \leq \beta \leq -0.25$. In the right-hand panel the red data points show the equivalent
information based on binning in stellar mass. The eight data points in
this plot correspond to bins of width 0.25 dex, spanning (left-to-right) the stellar-mass range $8.75 \leq \log(M_{\ast}/\Msun) \leq 10.75$.

The results shown in the left-hand panel of Fig.~7 provide
a clear illustration of the impact of scatter on the measured values
of UV-slope. Despite the fact that the input simulated galaxies follow
the IRX$-\beta$ relation expected for the \cite{calz00} attenuation
law (red curve), when the sample is binned by observed $\beta$, the binned values of IRX are significantly suppressed. 
Notably, for a value of $\sigma_{\beta}=0.35$, the output IRX$-\beta$
relation appears to follow the SMC-like IRX$-\beta$ relation with $\beta_{\rm int}=-2.62$ (green curve) determined by \cite{Reddy17}.

In contrast, the results shown in the right-hand panel of Fig.~7 show
the inherent advantages of binning by stellar mass. In this case, it
can be seen that the recovered IRX$-\beta$ relation closely follows the input IRX$-\beta$ relation.
In the simulation we have assumed a scatter of 0.3 dex in stellar mass, but the input IRX$-\beta$ relation is 
still recovered with reasonable accuracy for an assumed scatter as high as 0.5 dex.

\subsection{UV-slope uncertainty}
The results of our simplified simulation indicate that a typical
uncertainty of $\sigma_{\beta}\simeq 0.3$--0.4 is sufficient to bias an
input IRX$-\beta$ relation consistent with Calzetti-like attenuation, into an
observed IRX$-\beta$ relation which closely mimics that expected for
an SMC-like extinction law, combined with an ultra-blue value of
intrinsic UV slope. Consequently, it is of obvious interest to
determine whether or not this level of $\beta$ uncertainty is plausible.

In this study, and in \cite{Reddy17}, the UV spectral slope is measured
by fitting power laws to the available rest-frame UV
photometry. However, due to the need to exclude filters contaminated by Ly$\alpha$ emission 
and the requirement to restrict the long-wavelength anchor to
$\lambda_{\rm rest}<2580$\AA\, \citep{calz94}, it is often the case that there are only
two or three filters that sample the correct wavelength interval. 
This problem is particularly acute when dealing with the wide filters typically adopted for {\it HST} ACS imaging.
For example, for the purposes of our simulated sample at $z\simeq
2.5$, we have calculated synthetic photometry in the same {\it HST}
filters available in the HUDF and GOODS fields. In this case, the
only two filters available with which to obtain a clean measurement of
$\beta$ are  F606W ($V_{606}$) and F775W ($i_{775}$). 
In this situation it is straightforward to analytically determine the
relationship between the best-fitting power law and the $V_{606}-i_{775}$ colour:
\begin{equation}
\beta = 3.52\left( V_{606}-i_{775}\right)-2.0,
\end{equation}
where the multiplicative constant is determined by the pivot
wavelengths of the two filters. It can then be seen that the typical 
uncertainty on $\beta$ is given by
\begin{equation}
\sigma_{\beta} = 3.52\sqrt{\sigma_{606}^{2}+\sigma_{775}^{2}},
\end{equation}
where $\sigma_{606}$ and $\sigma_{775}$ are the uncertainties on the
$V_{606}$ and $i_{775}$ magnitudes. If we further assume that the F606W and F775W imaging is of equal depth,
equation (20) implies that the value of $\sigma_{\beta}=0.35$
employed in our simulation is equivalent to assuming $\simeq 15\sigma$ detections in both
filters. It would therefore appear reasonable to assume
that most studies of the IRX$-\beta$ relation at this epoch are subject to 
uncertainties of this level, or greater.


\subsection{Summary}
Although the simulation results we have presented here are  somewhat
simplified, they are sufficient to illustrate the general point that a
relatively small amount of uncertainty in measuring $\beta$ is
sufficient to produce biased results in the IRX$-\beta$ plane, when binning by $\beta$.
Although clearly not definitive, this does offer a plausible
explanation for why many IRX$-\beta$ results in the literature do not
follow the expected relation for Calzetti-like dust attenuation
whereas, in general, the IRX$-M_{\ast}$ results do.

\section{Conclusions}
In this paper we have presented the results of a study of the
IRX$-\beta$ and IRX$-M_{\ast}$ relations for star-forming galaxies at
$2<z<3$, based on the deep 1.3-mm ALMA mosaic of the HUDF presented by \cite{d17}.
In addition to the new ALMA-based results, we have determined the empirical relationship between UV slope and stellar mass for a large,
mass-complete, sample of $2<z<3$ star-forming galaxies selected over
an area of $\simeq 1$ deg$^{2}$. Armed with this information, and a compilation
of literature results, we have investigated the relationship between
stellar mass and UV attenuation in star-forming galaxies at this
crucial epoch. Our main conclusions can be summarised as follows:
\begin{enumerate}
\item{A stacking analysis within the 1.3-mm ALMA mosaic of the HUDF indicates
that star-forming galaxies at $2~<~z~<~3$  follow an IRX$-\beta$ relation
fully consistent with a relatively grey, Calzetti-like, dust attenuation law. In contrast, our new results are inconsistent with a dust law
    as steep in the UV as the SMC extinction law.} 

\item{Based on a large, mass complete, sample of star-forming galaxies
    at $2<z<3$, we find that UV slope is a smoothly
    increasing function of stellar mass. Our determination of the
    $\beta-M_{\ast}$ relation is well described by a third-order
    polynomial: $\beta = -1.136 +0.589X + 0.130X^{2} +
    0.106X^{3}$; where $X=\log(M_{\ast}/10^{10}\Msun)$. Furthermore,
    our SED modelling of the star-forming galaxy sample indicates that
    the intrinsic UV slope is fully consistent with a value of
    $\beta_{\rm int}\simeq -2.3$ over the full $8.50< \log(M_{\ast}/\Msun) < 11.5$ stellar-mass range.}

\item{Armed with the new empirical determination of the
    $\beta-M_{\ast}$ relation, it is possible to derive the expected
    relationship between $A_{1600}$ and stellar mass for any assumed dust law. By
    combining our new ALMA-based results with previous literature
    studies, we demonstrate that the $A_{1600}-M_{\ast}$ relation at
    $z\simeq 2.5$ is consistent with Calzetti-like attenuation at $\log(M_{\ast}/\Msun)\geq 9.75$.}

\item{By combining the observed $\beta$--$M_{\ast}$ relation with an
    assumed dust law, it is also possible to predict the form of the
    IRX$-M_{\ast}$ relation. We show that our new ALMA-based results,
    and the majority of recent literature results, are also consistent with
    the IRX-$M_{\ast}$ relation expected for Calzetti-like dust
    attenuation at $\log(M_{\ast}/\Msun)\geq 9.75$.}

\item{We find that the currently available IRX$-M_{\ast}$ data at $z\simeq 2.5$
    can be well described by a relationship of the form: $\log(\rm
    IRX)~=~0.85(\pm
    0.05)\log\left(\frac{M_{\ast}}{10^{10}\Msun}\right) + 0.99(\pm
    0.03)$, with an associated $1\sigma$ scatter of $\pm 0.17$ dex.}

\item{Our results indicate that an IRX$-M_{\ast}$ relation based on an SMC-like dust law,
    irrespective of the assumed value of the intrinsic UV slope ($\beta_{\rm int}$), has the wrong shape to adequately reproduce the observed IRX data over
    a large dynamic range in stellar mass. At stellar masses of $\log(M_{\ast}/\Msun)\leq 9.75$ the situation is much more
    uncertain, and deeper continuum imaging
    with ALMA will be required to determine
    the relationship between UV attenuation and stellar mass.}

\item{We employ a simple simulation to demonstrate that 
the uncertainty in individual $\beta$ measurements, combined with binning-up
samples by $\beta$, is sufficient to significantly bias measurements of
the IRX$-\beta$ relation. Crucially, our simulation also demonstrates that this bias does not
significantly affect determinations of the IRX$-\beta$ relation based on binning
samples by stellar mass.}

\item{The results of our simulation provide a plausible
    mechanism to explain the apparently contradictory IRX$-\beta$ and
    IRX$-M_{\ast}$ results in the literature, and demonstrate that
    both sets of observational results are fully consistent with a
    relatively grey, Calzetti-like, dust attenuation law.}

\item{The combination of observational and simulated results strongly
    indicate that, based on typical measurement uncertainties, it is stellar mass, rather than UV spectral slope,
    that provides the most accurate, least biased, proxy for the absolute
    level of UV attenuation experienced by star-forming galaxies.}
\end{enumerate}

\section*{Acknowledgements}
RJM, JSD, FC, NB, PNB and SK acknowledge the support of the UK Science and Technology
Facilities Council. RAAB acknowledges the support of the Oxford Centre
for Astrophysical Surveys which is funded through generous support
from the Hintze Family Charitable Foundation. JEG thanks the Royal Society for support.
MJM acknowledges the support of  the National Science Centre, Poland
through the POLONEZ grant 2015/19/P/ST9/04010 and the UK Science and
Technology Facilities Council. This project has received funding from
the European Union's Horizon 2020 research and innovation programme
under the Marie Sk{\l}odowska-Curie grant agreement No. 665778.
WR is supported by JSPS KAKENHI Grant Number JP15K17604 and the
Thailand Research Fund/Office of the Higher Education Commission Grant
Number MRG6080294. ALMA is a partnership of ESO (representing its member states), NSF (USA) and NINS (Japan), together
with NRC (Canada), and ASIAA (Taiwan), and KASI (Republic of Korea), in cooperation with the Republic
of Chile. The Joint ALMA Observatory is operated by ESO, AUI/NRAO and NAOJ.
This work is based in part on observations made with the NASA/ESA {\it Hubble Space Telescope}, which is operated by the Association 
of Universities for Research in Astronomy, Inc; under NASA contract NAS5-26555.
This work is also based in part on observations made with the {\it Spitzer Space Telescope}, which is operated by the Jet Propulsion Laboratory, 
California Institute of Technology under NASA contract 1407. {\it Herschel} is an ESA space observatory with science instruments provided by European-led
Principal Investigator consortia and with important participation from NASA.




\bibliographystyle{mnras}
\bibliography{rjm_irx_mnras_accepted} 

\begin{thebibliography}{}
\makeatletter
\relax
\def\mn@urlcharsother{\let\do\@makeother \do\$\do\&\do\#\do\^\do\_\do\%\do\~}
\def\mn@doi{\begingroup\mn@urlcharsother \@ifnextchar [ {\mn@doi@}
  {\mn@doi@[]}}
\def\mn@doi@[#1]#2{\def\@tempa{#1}\ifx\@tempa\@empty \href
  {http://dx.doi.org/#2} {doi:#2}\else \href {http://dx.doi.org/#2} {#1}\fi
  \endgroup}
\def\mn@eprint#1#2{\mn@eprint@#1:#2::\@nil}
\def\mn@eprint@arXiv#1{\href {http://arxiv.org/abs/#1} {{\tt arXiv:#1}}}
\def\mn@eprint@dblp#1{\href {http://dblp.uni-trier.de/rec/bibtex/#1.xml}
  {dblp:#1}}
\def\mn@eprint@#1:#2:#3:#4\@nil{\def\@tempa {#1}\def\@tempb {#2}\def\@tempc
  {#3}\ifx \@tempc \@empty \let \@tempc \@tempb \let \@tempb \@tempa \fi \ifx
  \@tempb \@empty \def\@tempb {arXiv}\fi \@ifundefined
  {mn@eprint@\@tempb}{\@tempb:\@tempc}{\expandafter \expandafter \csname
  mn@eprint@\@tempb\endcsname \expandafter{\@tempc}}}

\bibitem[\protect\citeauthoryear{{{\'A}lvarez-M{\'a}rquez}
  et~al.,}{{{\'A}lvarez-M{\'a}rquez} et~al.}{2016}]{am16}
{{\'A}lvarez-M{\'a}rquez} J.,  et~al., 2016, \mn@doi [\aap]
  {10.1051/0004-6361/201527190}, \href
  {http://adsabs.harvard.edu/abs/2016A%26A...587A.122A} {587, A122}

\bibitem[\protect\citeauthoryear{{Ashby} et~al.,}{{Ashby} et~al.}{2013}]{ash13}
{Ashby} M.~L.~N.,  et~al., 2013, \mn@doi [\apj] {10.1088/0004-637X/769/1/80},
  \href {http://adsabs.harvard.edu/abs/2013ApJ...769...80A} {769, 80}

\bibitem[\protect\citeauthoryear{{Asplund}, {Grevesse}, {Sauval}  \&
  {Scott}}{{Asplund} et~al.}{2009}]{asp09}
{Asplund} M.,  {Grevesse} N.,  {Sauval} A.~J.,   {Scott} P.,  2009, \mn@doi
  [\araa] {10.1146/annurev.astro.46.060407.145222}, \href
  {http://adsabs.harvard.edu/abs/2009ARA%26A..47..481A} {47, 481}

\bibitem[\protect\citeauthoryear{{Beckwith} et~al.,}{{Beckwith}
  et~al.}{2006}]{beck06}
{Beckwith} S.~V.~W.,  et~al., 2006, \mn@doi [\aj] {10.1086/507302}, \href
  {http://adsabs.harvard.edu/abs/2006AJ....132.1729B} {132, 1729}

\bibitem[\protect\citeauthoryear{{Behroozi}, {Wechsler}  \&
  {Conroy}}{{Behroozi} et~al.}{2013}]{beh13}
{Behroozi} P.~S.,  {Wechsler} R.~H.,   {Conroy} C.,  2013, \mn@doi [\apj]
  {10.1088/0004-637X/770/1/57}, \href
  {http://adsabs.harvard.edu/abs/2013ApJ...770...57B} {770, 57}

\bibitem[\protect\citeauthoryear{{Bourne} et~al.,}{{Bourne}
  et~al.}{2017}]{nathan17}
{Bourne} N.,  et~al., 2017, \mn@doi [\mnras] {10.1093/mnras/stx031}, \href
  {http://adsabs.harvard.edu/abs/2017MNRAS.tmp...41B} {}

\bibitem[\protect\citeauthoryear{{Bouwens} et~al.,}{{Bouwens}
  et~al.}{2011}]{bow11}
{Bouwens} R.~J.,  et~al., 2011, \mn@doi [\apj] {10.1088/0004-637X/737/2/90},
  \href {http://adsabs.harvard.edu/abs/2011ApJ...737...90B} {737, 90}

\bibitem[\protect\citeauthoryear{{Bouwens} et~al.,}{{Bouwens}
  et~al.}{2012}]{bow12}
{Bouwens} R.~J.,  et~al., 2012, \mn@doi [\apj] {10.1088/0004-637X/754/2/83},
  \href {http://adsabs.harvard.edu/abs/2012ApJ...754...83B} {754, 83}

\bibitem[\protect\citeauthoryear{{Bouwens} et~al.,}{{Bouwens}
  et~al.}{2014}]{bow14}
{Bouwens} R.~J.,  et~al., 2014, \mn@doi [\apj] {10.1088/0004-637X/793/2/115},
  \href {http://adsabs.harvard.edu/abs/2014ApJ...793..115B} {793, 115}

\bibitem[\protect\citeauthoryear{{Bouwens} et~al.,}{{Bouwens}
  et~al.}{2016}]{bow16}
{Bouwens} R.~J.,  et~al., 2016, \mn@doi [\apj] {10.3847/1538-4357/833/1/72},
  \href {http://adsabs.harvard.edu/abs/2016ApJ...833...72B} {833, 72}

\bibitem[\protect\citeauthoryear{{Brammer} et~al.,}{{Brammer}
  et~al.}{2012}]{brammer12}
{Brammer} G.~B.,  et~al., 2012, \mn@doi [\apjs] {10.1088/0067-0049/200/2/13},
  \href {http://adsabs.harvard.edu/abs/2012ApJS..200...13B} {200, 13}

\bibitem[\protect\citeauthoryear{{Bruzual} \& {Charlot}}{{Bruzual} \&
  {Charlot}}{2003}]{bc03}
{Bruzual} G.,  {Charlot} S.,  2003, \mn@doi [\mnras]
  {10.1046/j.1365-8711.2003.06897.x}, \href
  {http://adsabs.harvard.edu/abs/2003MNRAS.344.1000B} {344, 1000}

\bibitem[\protect\citeauthoryear{{Calzetti}}{{Calzetti}}{2001}]{calz01}
{Calzetti} D.,  2001, \mn@doi [\pasp] {10.1086/324269}, \href
  {http://adsabs.harvard.edu/abs/2001PASP..113.1449C} {113, 1449}

\bibitem[\protect\citeauthoryear{{Calzetti}, {Kinney}  \&
  {Storchi-Bergmann}}{{Calzetti} et~al.}{1994}]{calz94}
{Calzetti} D.,  {Kinney} A.~L.,   {Storchi-Bergmann} T.,  1994, \mn@doi [\apj]
  {10.1086/174346}, \href {http://adsabs.harvard.edu/abs/1994ApJ...429..582C}
  {429, 582}

\bibitem[\protect\citeauthoryear{{Calzetti}, {Armus}, {Bohlin}, {Kinney},
  {Koornneef}  \& {Storchi-Bergmann}}{{Calzetti} et~al.}{2000}]{calz00}
{Calzetti} D.,  {Armus} L.,  {Bohlin} R.~C.,  {Kinney} A.~L.,  {Koornneef} J.,
   {Storchi-Bergmann} T.,  2000, \mn@doi [\apj] {10.1086/308692}, \href
  {http://adsabs.harvard.edu/abs/2000ApJ...533..682C} {533, 682}

\bibitem[\protect\citeauthoryear{{Capak} et~al.,}{{Capak}
  et~al.}{2015}]{capak15}
{Capak} P.~L.,  et~al., 2015, \mn@doi [\nat] {10.1038/nature14500}, \href
  {http://adsabs.harvard.edu/abs/2015Natur.522..455C} {522, 455}

\bibitem[\protect\citeauthoryear{{Chabrier}}{{Chabrier}}{2003}]{chab03}
{Chabrier} G.,  2003, \mn@doi [\pasp] {10.1086/376392}, \href
  {http://adsabs.harvard.edu/abs/2003PASP..115..763C} {115, 763}

\bibitem[\protect\citeauthoryear{{Coppin} et~al.,}{{Coppin}
  et~al.}{2015}]{coppin15}
{Coppin} K.~E.~K.,  et~al., 2015, \mn@doi [\mnras] {10.1093/mnras/stu2185},
  \href {http://adsabs.harvard.edu/abs/2015MNRAS.446.1293C} {446, 1293}

\bibitem[\protect\citeauthoryear{{Cullen}, {Cirasuolo}, {Kewley}, {McLure},
  {Dunlop}  \& {Bowler}}{{Cullen} et~al.}{2016}]{ferg16}
{Cullen} F.,  {Cirasuolo} M.,  {Kewley} L.~J.,  {McLure} R.~J.,  {Dunlop}
  J.~S.,   {Bowler} R.~A.~A.,  2016, \mn@doi [\mnras] {10.1093/mnras/stw1181},
  \href {http://adsabs.harvard.edu/abs/2016MNRAS.460.3002C} {460, 3002}

\bibitem[\protect\citeauthoryear{{Cullen}, {McLure}, {Khochfar}, {Dunlop}  \&
  {Dalla Vecchia}}{{Cullen} et~al.}{2017}]{ferg17}
{Cullen} F.,  {McLure} R.~J.,  {Khochfar} S.,  {Dunlop} J.~S.,   {Dalla
  Vecchia} C.,  2017, \mn@doi [\mnras] {10.1093/mnras/stx1451}, \href
  {http://adsabs.harvard.edu/abs/2017MNRAS.470.3006C} {470, 3006}

\bibitem[\protect\citeauthoryear{{Dahlen} et~al.,}{{Dahlen}
  et~al.}{2013}]{dahlen13}
{Dahlen} T.,  et~al., 2013, \mn@doi [\apj] {10.1088/0004-637X/775/2/93}, \href
  {http://adsabs.harvard.edu/abs/2013ApJ...775...93D} {775, 93}

\bibitem[\protect\citeauthoryear{{Davidzon} et~al.,}{{Davidzon}
  et~al.}{2017}]{david17}
{Davidzon} I.,  et~al., 2017, \mn@doi [\aap] {10.1051/0004-6361/201730419},
  \href {http://adsabs.harvard.edu/abs/2017A%26A...605A..70D} {605, A70}

\bibitem[\protect\citeauthoryear{{Dunlop} et~al.,}{{Dunlop} et~al.}{2013}]{d13}
{Dunlop} J.~S.,  et~al., 2013, \mn@doi [\mnras] {10.1093/mnras/stt702}, \href
  {http://adsabs.harvard.edu/abs/2013MNRAS.432.3520D} {432, 3520}

\bibitem[\protect\citeauthoryear{{Dunlop} et~al.,}{{Dunlop} et~al.}{2017}]{d17}
{Dunlop} J.~S.,  et~al., 2017, \mn@doi [\mnras] {10.1093/mnras/stw3088}, \href
  {http://adsabs.harvard.edu/abs/2017MNRAS.466..861D} {466, 861}

\bibitem[\protect\citeauthoryear{{Eldridge} \& {Stanway}}{{Eldridge} \&
  {Stanway}}{2016}]{jj16}
{Eldridge} J.~J.,  {Stanway} E.~R.,  2016, \mn@doi [\mnras]
  {10.1093/mnras/stw1772}, \href
  {http://adsabs.harvard.edu/abs/2016MNRAS.462.3302E} {462, 3302}

\bibitem[\protect\citeauthoryear{{Ellis} et~al.,}{{Ellis}
  et~al.}{2013}]{ellis13}
{Ellis} R.~S.,  et~al., 2013, \mn@doi [\apjl] {10.1088/2041-8205/763/1/L7},
  \href {http://adsabs.harvard.edu/abs/2013ApJ...763L...7E} {763, L7}

\bibitem[\protect\citeauthoryear{{Ferland} et~al.,}{{Ferland}
  et~al.}{2013}]{fer13}
{Ferland} G.~J.,  et~al., 2013, \rmxaa, \href
  {http://adsabs.harvard.edu/abs/2013RMxAA..49..137F} {49, 137}

\bibitem[\protect\citeauthoryear{{Finkelstein}}{{Finkelstein}}{2016}]{fink16}
{Finkelstein} S.~L.,  2016, \mn@doi [\pasa] {10.1017/pasa.2016.26}, \href
  {http://adsabs.harvard.edu/abs/2016PASA...33...37F} {33, e037}

\bibitem[\protect\citeauthoryear{{Finkelstein} et~al.,}{{Finkelstein}
  et~al.}{2012}]{fink12}
{Finkelstein} S.~L.,  et~al., 2012, \mn@doi [\apj]
  {10.1088/0004-637X/756/2/164}, \href
  {http://adsabs.harvard.edu/abs/2012ApJ...756..164F} {756, 164}

\bibitem[\protect\citeauthoryear{{Fontana} et~al.,}{{Fontana}
  et~al.}{2014}]{font14}
{Fontana} A.,  et~al., 2014, \mn@doi [\aap] {10.1051/0004-6361/201423543},
  \href {http://adsabs.harvard.edu/abs/2014A} {570, A11}

\bibitem[\protect\citeauthoryear{{Fudamoto} et~al.,}{{Fudamoto}
  et~al.}{2017}]{fuda17}
{Fudamoto} Y.,  et~al., 2017, \mn@doi [\mnras] {10.1093/mnras/stx1948}, \href
  {http://adsabs.harvard.edu/abs/2017MNRAS.472..483F} {472, 483}

\bibitem[\protect\citeauthoryear{{Furusawa} et~al.,}{{Furusawa}
  et~al.}{2016}]{furusawa16}
{Furusawa} H.,  et~al., 2016, \mn@doi [\apj] {10.3847/0004-637X/822/1/46},
  \href {http://adsabs.harvard.edu/abs/2016ApJ...822...46F} {822, 46}

\bibitem[\protect\citeauthoryear{{Galametz} et~al.,}{{Galametz}
  et~al.}{2013}]{gal13}
{Galametz} A.,  et~al., 2013, \mn@doi [\apjs] {10.1088/0067-0049/206/2/10},
  \href {http://adsabs.harvard.edu/abs/2013ApJS..206...10G} {206, 10}

\bibitem[\protect\citeauthoryear{{Garn} \& {Best}}{{Garn} \&
  {Best}}{2010}]{garn10}
{Garn} T.,  {Best} P.~N.,  2010, \mn@doi [\mnras]
  {10.1111/j.1365-2966.2010.17321.x}, \href
  {http://adsabs.harvard.edu/abs/2010MNRAS.409..421G} {409, 421}

\bibitem[\protect\citeauthoryear{{Geach} et~al.,}{{Geach} et~al.}{2017}]{cls}
{Geach} J.~E.,  et~al., 2017, \mn@doi [\mnras] {10.1093/mnras/stw2721}, \href
  {http://adsabs.harvard.edu/abs/2017MNRAS.465.1789G} {465, 1789}

\bibitem[\protect\citeauthoryear{{Gordon}, {Clayton}, {Misselt}, {Landolt}  \&
  {Wolff}}{{Gordon} et~al.}{2003}]{gord03}
{Gordon} K.~D.,  {Clayton} G.~C.,  {Misselt} K.~A.,  {Landolt} A.~U.,   {Wolff}
  M.~J.,  2003, \mn@doi [\apj] {10.1086/376774}, \href
  {http://adsabs.harvard.edu/abs/2003ApJ...594..279G} {594, 279}

\bibitem[\protect\citeauthoryear{{Guo} et~al.,}{{Guo} et~al.}{2013}]{guo13}
{Guo} Y.,  et~al., 2013, \mn@doi [\apjs] {10.1088/0067-0049/207/2/24}, \href
  {http://adsabs.harvard.edu/abs/2013ApJS..207...24G} {207, 24}

\bibitem[\protect\citeauthoryear{{Heinis} et~al.,}{{Heinis}
  et~al.}{2013}]{heinis13}
{Heinis} S.,  et~al., 2013, \mn@doi [\mnras] {10.1093/mnras/sts397}, \href
  {http://adsabs.harvard.edu/abs/2013MNRAS.429.1113H} {429, 1113}

\bibitem[\protect\citeauthoryear{{Heinis} et~al.,}{{Heinis}
  et~al.}{2014}]{heinis14}
{Heinis} S.,  et~al., 2014, \mn@doi [\mnras] {10.1093/mnras/stt1960}, \href
  {http://adsabs.harvard.edu/abs/2014MNRAS.437.1268H} {437, 1268}

\bibitem[\protect\citeauthoryear{{Helou}, {Khan}, {Malek}  \&
  {Boehmer}}{{Helou} et~al.}{1988}]{hel88}
{Helou} G.,  {Khan} I.~R.,  {Malek} L.,   {Boehmer} L.,  1988, \mn@doi [\apjs]
  {10.1086/191285}, \href {http://adsabs.harvard.edu/abs/1988ApJS...68..151H}
  {68, 151}

\bibitem[\protect\citeauthoryear{{Hudelot} et~al.,}{{Hudelot}
  et~al.}{2012}]{cfhtls}
{Hudelot} P.,  et~al., 2012, VizieR Online Data Catalog, \href
  {http://adsabs.harvard.edu/abs/2012yCat.2317....0H} {2317}

\bibitem[\protect\citeauthoryear{{Ibar} et~al.,}{{Ibar} et~al.}{2013}]{ibar13}
{Ibar} E.,  et~al., 2013, \mn@doi [\mnras] {10.1093/mnras/stt1258}, \href
  {http://adsabs.harvard.edu/abs/2013MNRAS.434.3218I} {434, 3218}

\bibitem[\protect\citeauthoryear{{Illingworth} et~al.,}{{Illingworth}
  et~al.}{2013}]{ill13}
{Illingworth} G.~D.,  et~al., 2013, \mn@doi [\apjs]
  {10.1088/0067-0049/209/1/6}, \href
  {http://adsabs.harvard.edu/abs/2013ApJS..209....6I} {209, 6}

\bibitem[\protect\citeauthoryear{{Kashino} et~al.,}{{Kashino}
  et~al.}{2013}]{kash13}
{Kashino} D.,  et~al., 2013, \mn@doi [\apjl] {10.1088/2041-8205/777/1/L8},
  \href {http://adsabs.harvard.edu/abs/2013ApJ...777L...8K} {777, L8}

\bibitem[\protect\citeauthoryear{{Kennicutt} \& {Evans}}{{Kennicutt} \&
  {Evans}}{2012}]{ken12}
{Kennicutt} R.~C.,  {Evans} N.~J.,  2012, \mn@doi [\araa]
  {10.1146/annurev-astro-081811-125610}, \href
  {http://adsabs.harvard.edu/abs/2012ARA%26A..50..531K} {50, 531}

\bibitem[\protect\citeauthoryear{{Koekemoer} et~al.,}{{Koekemoer}
  et~al.}{2013}]{koke13}
{Koekemoer} A.~M.,  et~al., 2013, \mn@doi [\apjs] {10.1088/0067-0049/209/1/3},
  \href {http://adsabs.harvard.edu/abs/2013ApJS..209....3K} {209, 3}

\bibitem[\protect\citeauthoryear{{Kong}, {Charlot}, {Brinchmann}  \&
  {Fall}}{{Kong} et~al.}{2004}]{kong04}
{Kong} X.,  {Charlot} S.,  {Brinchmann} J.,   {Fall} S.~M.,  2004, \mn@doi
  [\mnras] {10.1111/j.1365-2966.2004.07556.x}, \href
  {http://adsabs.harvard.edu/abs/2004MNRAS.349..769K} {349, 769}

\bibitem[\protect\citeauthoryear{{Koprowski} et~al.,}{{Koprowski}
  et~al.}{2016}]{majec16}
{Koprowski} M.~P.,  et~al., 2016, \mn@doi [\apjl]
  {10.3847/2041-8205/828/2/L21}, \href
  {http://adsabs.harvard.edu/abs/2016ApJ...828L..21K} {828, L21}

\bibitem[\protect\citeauthoryear{{Kroupa}}{{Kroupa}}{2001}]{kroupa01}
{Kroupa} P.,  2001, \mn@doi [\mnras] {10.1046/j.1365-8711.2001.04022.x}, \href
  {http://adsabs.harvard.edu/abs/2001MNRAS.322..231K} {322, 231}

\bibitem[\protect\citeauthoryear{{Labb{\'e}} et~al.,}{{Labb{\'e}}
  et~al.}{2015}]{labbe15}
{Labb{\'e}} I.,  et~al., 2015, \mn@doi [\apjs] {10.1088/0067-0049/221/2/23},
  \href {http://adsabs.harvard.edu/abs/2015ApJS..221...23L} {221, 23}

\bibitem[\protect\citeauthoryear{{Laidler} et~al.,}{{Laidler}
  et~al.}{2007}]{laidler07}
{Laidler} V.~G.,  et~al., 2007, \mn@doi [\pasp] {10.1086/523898}, \href
  {http://adsabs.harvard.edu/abs/2007PASP..119.1325L} {119, 1325}

\bibitem[\protect\citeauthoryear{{Laporte} et~al.,}{{Laporte}
  et~al.}{2017}]{laporte17}
{Laporte} N.,  et~al., 2017, \mn@doi [\apjl] {10.3847/2041-8213/aa62aa}, \href
  {http://adsabs.harvard.edu/abs/2017ApJ...837L..21L} {837, L21}

\bibitem[\protect\citeauthoryear{{Lee}, {Alberts}, {Atlee}, {Dey}, {Pope},
  {Jannuzi}, {Reddy}  \& {Brown}}{{Lee} et~al.}{2012}]{lee12}
{Lee} K.-S.,  {Alberts} S.,  {Atlee} D.,  {Dey} A.,  {Pope} A.,  {Jannuzi}
  B.~T.,  {Reddy} N.,   {Brown} M.~J.~I.,  2012, \mn@doi [\apjl]
  {10.1088/2041-8205/758/2/L31}, \href
  {http://adsabs.harvard.edu/abs/2012ApJ...758L..31L} {758, L31}

\bibitem[\protect\citeauthoryear{{Madau}}{{Madau}}{1995}]{mad95}
{Madau} P.,  1995, \mn@doi [\apj] {10.1086/175332}, \href
  {http://adsabs.harvard.edu/abs/1995ApJ...441...18M} {441, 18}

\bibitem[\protect\citeauthoryear{{Madau} \& {Dickinson}}{{Madau} \&
  {Dickinson}}{2014}]{mad14}
{Madau} P.,  {Dickinson} M.,  2014, \mn@doi [\araa]
  {10.1146/annurev-astro-081811-125615}, \href
  {http://adsabs.harvard.edu/abs/2014ARA%26A..52..415M} {52, 415}

\bibitem[\protect\citeauthoryear{{McLeod}, {McLure}, {Dunlop}, {Robertson},
  {Ellis}  \& {Targett}}{{McLeod} et~al.}{2015}]{derek15}
{McLeod} D.~J.,  {McLure} R.~J.,  {Dunlop} J.~S.,  {Robertson} B.~E.,  {Ellis}
  R.~S.,   {Targett} T.~A.,  2015, \mn@doi [\mnras] {10.1093/mnras/stv780},
  \href {http://adsabs.harvard.edu/abs/2015MNRAS.450.3032M} {450, 3032}

\bibitem[\protect\citeauthoryear{{McLeod}, {McLure}  \& {Dunlop}}{{McLeod}
  et~al.}{2016}]{derek16}
{McLeod} D.~J.,  {McLure} R.~J.,   {Dunlop} J.~S.,  2016, \mn@doi [\mnras]
  {10.1093/mnras/stw904}, \href
  {http://adsabs.harvard.edu/abs/2016MNRAS.459.3812M} {459, 3812}

\bibitem[\protect\citeauthoryear{{McLure} et~al.,}{{McLure}
  et~al.}{2011}]{mclure11}
{McLure} R.~J.,  et~al., 2011, \mn@doi [\mnras]
  {10.1111/j.1365-2966.2011.19626.x}, \href
  {http://adsabs.harvard.edu/abs/2011MNRAS.418.2074M} {418, 2074}

\bibitem[\protect\citeauthoryear{{McLure} et~al.,}{{McLure}
  et~al.}{2013}]{mclure13}
{McLure} R.~J.,  et~al., 2013, \mn@doi [\mnras] {10.1093/mnras/stt627}, \href
  {http://adsabs.harvard.edu/abs/2013MNRAS.432.2696M} {432, 2696}

\bibitem[\protect\citeauthoryear{{Meurer}, {Heckman}  \& {Calzetti}}{{Meurer}
  et~al.}{1999}]{m99}
{Meurer} G.~R.,  {Heckman} T.~M.,   {Calzetti} D.,  1999, \mn@doi [\apj]
  {10.1086/307523}, \href {http://adsabs.harvard.edu/abs/1999ApJ...521...64M}
  {521, 64}

\bibitem[\protect\citeauthoryear{{Momcheva} et~al.,}{{Momcheva}
  et~al.}{2016}]{mom16}
{Momcheva} I.~G.,  et~al., 2016, \mn@doi [\apjs] {10.3847/0067-0049/225/2/27},
  \href {http://adsabs.harvard.edu/abs/2016ApJS..225...27M} {225, 27}

\bibitem[\protect\citeauthoryear{{Mortlock}, {McLure}, {Bowler}, {McLeod},
  {M{\'a}rmol-Queralt{\'o}}, {Parsa}, {Dunlop}  \& {Bruce}}{{Mortlock}
  et~al.}{2017}]{mort17}
{Mortlock} A.,  {McLure} R.~J.,  {Bowler} R.~A.~A.,  {McLeod} D.~J.,
  {M{\'a}rmol-Queralt{\'o}} E.,  {Parsa} S.,  {Dunlop} J.~S.,   {Bruce} V.~A.,
  2017, \mn@doi [\mnras] {10.1093/mnras/stw2728}, \href
  {http://adsabs.harvard.edu/abs/2017MNRAS.465..672M} {465, 672}

\bibitem[\protect\citeauthoryear{{Narayanan}, {Dav{\'e}}, {Johnson},
  {Thompson}, {Conroy}  \& {Geach}}{{Narayanan} et~al.}{2018}]{des17}
{Narayanan} D.,  {Dav{\'e}} R.,  {Johnson} B.~D.,  {Thompson} R.,  {Conroy} C.,
    {Geach} J.,  2018, \mn@doi [\mnras] {10.1093/mnras/stx2860}, \href
  {http://adsabs.harvard.edu/abs/2018MNRAS.474.1718N} {474, 1718}

\bibitem[\protect\citeauthoryear{{Nonino} et~al.,}{{Nonino}
  et~al.}{2009}]{non09}
{Nonino} M.,  et~al., 2009, \mn@doi [\apjs] {10.1088/0067-0049/183/2/244},
  \href {http://adsabs.harvard.edu/abs/2009ApJS..183..244N} {183, 244}

\bibitem[\protect\citeauthoryear{{Novak} et~al.,}{{Novak}
  et~al.}{2017}]{novak17}
{Novak} M.,  et~al., 2017, \mn@doi [\aap] {10.1051/0004-6361/201629436}, \href
  {http://adsabs.harvard.edu/abs/2017A%26A...602A...5N} {602, A5}

\bibitem[\protect\citeauthoryear{{Oke} \& {Gunn}}{{Oke} \&
  {Gunn}}{1983}]{oke83}
{Oke} J.~B.,  {Gunn} J.~E.,  1983, \mn@doi [\apj] {10.1086/160817}, \href
  {http://adsabs.harvard.edu/abs/1983ApJ...266..713O} {266, 713}

\bibitem[\protect\citeauthoryear{{Overzier} et~al.,}{{Overzier}
  et~al.}{2011}]{over11}
{Overzier} R.~A.,  et~al., 2011, \mn@doi [\apjl] {10.1088/2041-8205/726/1/L7},
  \href {http://adsabs.harvard.edu/abs/2011ApJ...726L...7O} {726, L7}

\bibitem[\protect\citeauthoryear{{Paardekooper}, {Khochfar}  \& {Dalla
  Vecchia}}{{Paardekooper} et~al.}{2015}]{koop15}
{Paardekooper} J.-P.,  {Khochfar} S.,   {Dalla Vecchia} C.,  2015, \mn@doi
  [\mnras] {10.1093/mnras/stv1114}, \href
  {http://adsabs.harvard.edu/abs/2015MNRAS.451.2544P} {451, 2544}

\bibitem[\protect\citeauthoryear{{Pannella} et~al.,}{{Pannella}
  et~al.}{2015}]{pan15}
{Pannella} M.,  et~al., 2015, \mn@doi [\apj] {10.1088/0004-637X/807/2/141},
  \href {http://adsabs.harvard.edu/abs/2015ApJ...807..141P} {807, 141}

\bibitem[\protect\citeauthoryear{{Pettini}, {Kellogg}, {Steidel}, {Dickinson},
  {Adelberger}  \& {Giavalisco}}{{Pettini} et~al.}{1998}]{pet98}
{Pettini} M.,  {Kellogg} M.,  {Steidel} C.~C.,  {Dickinson} M.,  {Adelberger}
  K.~L.,   {Giavalisco} M.,  1998, \mn@doi [\apj] {10.1086/306431}, \href
  {http://adsabs.harvard.edu/abs/1998ApJ...508..539P} {508, 539}

\bibitem[\protect\citeauthoryear{{Pope} et~al.,}{{Pope} et~al.}{2017}]{pope17}
{Pope} A.,  et~al., 2017, \mn@doi [\apj] {10.3847/1538-4357/aa6573}, \href
  {http://adsabs.harvard.edu/abs/2017ApJ...838..137P} {838, 137}

\bibitem[\protect\citeauthoryear{{Popping}, {Puglisi}  \& {Norman}}{{Popping}
  et~al.}{2017}]{popping17}
{Popping} G.,  {Puglisi} A.,   {Norman} C.~A.,  2017, \mn@doi [\mnras]
  {10.1093/mnras/stx2202}, \href
  {http://adsabs.harvard.edu/abs/2017MNRAS.472.2315P} {472, 2315}

\bibitem[\protect\citeauthoryear{{Prevot}, {Lequeux}, {Prevot}, {Maurice}  \&
  {Rocca-Volmerange}}{{Prevot} et~al.}{1984}]{prev84}
{Prevot} M.~L.,  {Lequeux} J.,  {Prevot} L.,  {Maurice} E.,
  {Rocca-Volmerange} B.,  1984, \aap, \href
  {http://adsabs.harvard.edu/abs/1984A%26A...132..389P} {132, 389}

\bibitem[\protect\citeauthoryear{{Price} et~al.,}{{Price}
  et~al.}{2014}]{price14}
{Price} S.~H.,  et~al., 2014, \mn@doi [\apj] {10.1088/0004-637X/788/1/86},
  \href {http://adsabs.harvard.edu/abs/2014ApJ...788...86P} {788, 86}

\bibitem[\protect\citeauthoryear{{Reddy}, {Erb}, {Pettini}, {Steidel}  \&
  {Shapley}}{{Reddy} et~al.}{2010}]{red10}
{Reddy} N.~A.,  {Erb} D.~K.,  {Pettini} M.,  {Steidel} C.~C.,   {Shapley}
  A.~E.,  2010, \mn@doi [\apj] {10.1088/0004-637X/712/2/1070}, \href
  {http://adsabs.harvard.edu/abs/2010ApJ...712.1070R} {712, 1070}

\bibitem[\protect\citeauthoryear{{Reddy} et~al.,}{{Reddy} et~al.}{2015}]{red15}
{Reddy} N.~A.,  et~al., 2015, \mn@doi [\apj] {10.1088/0004-637X/806/2/259},
  \href {http://adsabs.harvard.edu/abs/2015ApJ...806..259R} {806, 259}

\bibitem[\protect\citeauthoryear{{Reddy} et~al.,}{{Reddy}
  et~al.}{2018}]{Reddy17}
{Reddy} N.~A.,  et~al., 2018, \mn@doi [\apj] {10.3847/1538-4357/aaa3e7}, \href
  {http://adsabs.harvard.edu/abs/2018ApJ...853...56R} {853, 56}

\bibitem[\protect\citeauthoryear{{Rogers}, {McLure}  \& {Dunlop}}{{Rogers}
  et~al.}{2013}]{sandy13}
{Rogers} A.~B.,  {McLure} R.~J.,   {Dunlop} J.~S.,  2013, \mn@doi [\mnras]
  {10.1093/mnras/sts515}, \href
  {http://adsabs.harvard.edu/abs/2013MNRAS.429.2456R} {429, 2456}

\bibitem[\protect\citeauthoryear{{Rogers} et~al.,}{{Rogers}
  et~al.}{2014}]{sandy14}
{Rogers} A.~B.,  et~al., 2014, \mn@doi [\mnras] {10.1093/mnras/stu558}, \href
  {http://adsabs.harvard.edu/abs/2014MNRAS.440.3714R} {440, 3714}

\bibitem[\protect\citeauthoryear{{Rujopakarn} et~al.,}{{Rujopakarn}
  et~al.}{2016}]{whipu16}
{Rujopakarn} W.,  et~al., 2016, \mn@doi [\apj] {10.3847/0004-637X/833/1/12},
  \href {http://adsabs.harvard.edu/abs/2016ApJ...833...12R} {833, 12}

\bibitem[\protect\citeauthoryear{{Salmon} et~al.,}{{Salmon}
  et~al.}{2016}]{salmon16}
{Salmon} B.,  et~al., 2016, \mn@doi [\apj] {10.3847/0004-637X/827/1/20}, \href
  {http://adsabs.harvard.edu/abs/2016ApJ...827...20S} {827, 20}

\bibitem[\protect\citeauthoryear{{Salpeter}}{{Salpeter}}{1955}]{salp55}
{Salpeter} E.~E.,  1955, \mn@doi [\apj] {10.1086/145971}, \href
  {http://adsabs.harvard.edu/abs/1955ApJ...121..161S} {121, 161}

\bibitem[\protect\citeauthoryear{{Santini} et~al.,}{{Santini}
  et~al.}{2015}]{sant15}
{Santini} P.,  et~al., 2015, \mn@doi [\apj] {10.1088/0004-637X/801/2/97}, \href
  {http://adsabs.harvard.edu/abs/2015ApJ...801...97S} {801, 97}

\bibitem[\protect\citeauthoryear{{Scoville}, {Faisst}, {Capak}, {Kakazu}, {Li}
  \& {Steinhardt}}{{Scoville} et~al.}{2015}]{scov15}
{Scoville} N.,  {Faisst} A.,  {Capak} P.,  {Kakazu} Y.,  {Li} G.,
  {Steinhardt} C.,  2015, \mn@doi [\apj] {10.1088/0004-637X/800/2/108}, \href
  {http://adsabs.harvard.edu/abs/2015ApJ...800..108S} {800, 108}

\bibitem[\protect\citeauthoryear{{Seon} \& {Draine}}{{Seon} \&
  {Draine}}{2016}]{seon16}
{Seon} K.-I.,  {Draine} B.~T.,  2016, \mn@doi [\apj]
  {10.3847/1538-4357/833/2/201}, \href
  {http://adsabs.harvard.edu/abs/2016ApJ...833..201S} {833, 201}

\bibitem[\protect\citeauthoryear{{Shapley} et~al.,}{{Shapley}
  et~al.}{2015}]{shap15}
{Shapley} A.~E.,  et~al., 2015, \mn@doi [\apj] {10.1088/0004-637X/801/2/88},
  \href {http://adsabs.harvard.edu/abs/2015ApJ...801...88S} {801, 88}

\bibitem[\protect\citeauthoryear{{Speagle}, {Steinhardt}, {Capak}  \&
  {Silverman}}{{Speagle} et~al.}{2014}]{speagle14}
{Speagle} J.~S.,  {Steinhardt} C.~L.,  {Capak} P.~L.,   {Silverman} J.~D.,
  2014, \mn@doi [\apjs] {10.1088/0067-0049/214/2/15}, \href
  {http://adsabs.harvard.edu/abs/2014ApJS..214...15S} {214, 15}

\bibitem[\protect\citeauthoryear{{Stanway}, {Eldridge}  \& {Becker}}{{Stanway}
  et~al.}{2016}]{stan16}
{Stanway} E.~R.,  {Eldridge} J.~J.,   {Becker} G.~D.,  2016, \mn@doi [\mnras]
  {10.1093/mnras/stv2661}, \href
  {http://adsabs.harvard.edu/abs/2016MNRAS.456..485S} {456, 485}

\bibitem[\protect\citeauthoryear{{Stark}}{{Stark}}{2016}]{stark16}
{Stark} D.~P.,  2016, \mn@doi [\araa] {10.1146/annurev-astro-081915-023417},
  \href {http://adsabs.harvard.edu/abs/2016ARA%26A..54..761S} {54, 761}

\bibitem[\protect\citeauthoryear{{Steidel}, {Strom}, {Pettini}, {Rudie},
  {Reddy}  \& {Trainor}}{{Steidel} et~al.}{2016}]{steidel16}
{Steidel} C.~C.,  {Strom} A.~L.,  {Pettini} M.,  {Rudie} G.~C.,  {Reddy} N.~A.,
    {Trainor} R.~F.,  2016, \mn@doi [\apj] {10.3847/0004-637X/826/2/159}, \href
  {http://adsabs.harvard.edu/abs/2016ApJ...826..159S} {826, 159}

\bibitem[\protect\citeauthoryear{{Steinhardt} et~al.,}{{Steinhardt}
  et~al.}{2014}]{splash}
{Steinhardt} C.~L.,  et~al., 2014, \mn@doi [\apjl]
  {10.1088/2041-8205/791/2/L25}, \href
  {http://adsabs.harvard.edu/abs/2014ApJ...791L..25S} {791, L25}

\bibitem[\protect\citeauthoryear{{Strom}, {Steidel}, {Rudie}, {Trainor},
  {Pettini}  \& {Reddy}}{{Strom} et~al.}{2017}]{strom17}
{Strom} A.~L.,  {Steidel} C.~C.,  {Rudie} G.~C.,  {Trainor} R.~F.,  {Pettini}
  M.,   {Reddy} N.~A.,  2017, \mn@doi [\apj] {10.3847/1538-4357/836/2/164},
  \href {http://adsabs.harvard.edu/abs/2017ApJ...836..164S} {836, 164}

\bibitem[\protect\citeauthoryear{{Takeuchi}, {Yuan}, {Ikeyama}, {Murata}  \&
  {Inoue}}{{Takeuchi} et~al.}{2012}]{take12}
{Takeuchi} T.~T.,  {Yuan} F.-T.,  {Ikeyama} A.,  {Murata} K.~L.,   {Inoue}
  A.~K.,  2012, \mn@doi [\apj] {10.1088/0004-637X/755/2/144}, \href
  {http://adsabs.harvard.edu/abs/2012ApJ...755..144T} {755, 144}

\bibitem[\protect\citeauthoryear{{Tomczak} et~al.,}{{Tomczak}
  et~al.}{2014}]{tom14}
{Tomczak} A.~R.,  et~al., 2014, \mn@doi [\apj] {10.1088/0004-637X/783/2/85},
  \href {http://adsabs.harvard.edu/abs/2014ApJ...783...85T} {783, 85}

\bibitem[\protect\citeauthoryear{{Watson}, {Christensen}, {Knudsen}, {Richard},
  {Gallazzi}  \& {Micha{\l}owski}}{{Watson} et~al.}{2015}]{watson15}
{Watson} D.,  {Christensen} L.,  {Knudsen} K.~K.,  {Richard} J.,  {Gallazzi}
  A.,   {Micha{\l}owski} M.~J.,  2015, \mn@doi [\nat] {10.1038/nature14164},
  \href {http://adsabs.harvard.edu/abs/2015Natur.519..327W} {519, 327}

\bibitem[\protect\citeauthoryear{{Whitaker} et~al.,}{{Whitaker}
  et~al.}{2011}]{whit11}
{Whitaker} K.~E.,  et~al., 2011, \mn@doi [\apj] {10.1088/0004-637X/735/2/86},
  \href {http://adsabs.harvard.edu/abs/2011ApJ...735...86W} {735, 86}

\bibitem[\protect\citeauthoryear{{Whitaker} et~al.,}{{Whitaker}
  et~al.}{2014}]{whit14}
{Whitaker} K.~E.,  et~al., 2014, \mn@doi [\apj] {10.1088/0004-637X/795/2/104},
  \href {http://adsabs.harvard.edu/abs/2014ApJ...795..104W} {795, 104}

\bibitem[\protect\citeauthoryear{{Williams}, {Quadri}, {Franx}, {van Dokkum}
  \& {Labb{\'e}}}{{Williams} et~al.}{2009}]{will09}
{Williams} R.~J.,  {Quadri} R.~F.,  {Franx} M.,  {van Dokkum} P.,   {Labb{\'e}}
  I.,  2009, \mn@doi [\apj] {10.1088/0004-637X/691/2/1879}, \href
  {http://adsabs.harvard.edu/abs/2009ApJ...691.1879W} {691, 1879}

\bibitem[\protect\citeauthoryear{{Witt} \& {Gordon}}{{Witt} \&
  {Gordon}}{2000}]{witt2000}
{Witt} A.~N.,  {Gordon} K.~D.,  2000, \mn@doi [\apj] {10.1086/308197}, \href
  {http://adsabs.harvard.edu/abs/2000ApJ...528..799W} {528, 799}

\bibitem[\protect\citeauthoryear{{Wuyts} et~al.,}{{Wuyts}
  et~al.}{2013}]{wuyts13}
{Wuyts} S.,  et~al., 2013, \mn@doi [\apj] {10.1088/0004-637X/779/2/135}, \href
  {http://adsabs.harvard.edu/abs/2013ApJ...779..135W} {779, 135}

\bibitem[\protect\citeauthoryear{{Zafar}, {Watson}, {Fynbo}, {Malesani},
  {Jakobsson}  \& {de Ugarte Postigo}}{{Zafar} et~al.}{2011}]{grb}
{Zafar} T.,  {Watson} D.,  {Fynbo} J.~P.~U.,  {Malesani} D.,  {Jakobsson} P.,
  {de Ugarte Postigo} A.,  2011, \mn@doi [\aap] {10.1051/0004-6361/201116663},
  \href {http://adsabs.harvard.edu/abs/2011A%26A...532A.143Z} {532, A143}

\makeatother
\end{thebibliography}

\bsp	
\label{lastpage}
\end{document}